\documentclass[preprint]{revtex4-1}

\usepackage{textcomp}

\usepackage{amssymb}
\usepackage{graphicx}
\usepackage{subfigure}
\usepackage{amsfonts}
\usepackage{amssymb}
\usepackage{latexsym}
\usepackage{natbib}
\usepackage{bm}
\usepackage{newlfont}
\usepackage{epsfig}
\usepackage{ctable}
\usepackage{amsmath}
\usepackage[american]{babel}

\usepackage{placeins}
\usepackage{cleveref}

\usepackage{lineno}

\crefname{section}{§}{§§}
\Crefname{section}{§}{§§}

\begin{document}

\title{Stick-slip-to-stick transition of liquid oscillation in a U-tube}

\author{A.~Bongarzone and F.~Gallaire  }
\affiliation{Laboratory of Fluid Mechanics and Instabilities, \'Ecole Polytechnique F\'ed\'erale de Lausanne, Lausanne, CH-1015, Switzerland}

\begin{abstract}
The nonlinear decay of oscillations of a liquid column in a U-shaped tube is investigated within the theoretical framework of the projection method formalized by Bongarzone \textit{et al.} (2021) \citep{bongarzone2021relaxation}. Starting from the full hydrodynamic system supplemented by a phenomenological contact line model, this physics-inspired method uses successive linear eigenmode projections to simulate the relaxation dynamics of liquid oscillations in the presence of sliding triple lines. Each projection is shown to eventually induce a rapid loss of total energy in the liquid motion, thus contributing to its nonlinear damping. A thorough quantitative comparison with experiments by Dollet \textit{et al.} (2020) \citep{dollet2020transition} demonstrates that, in contradistinction with their simplistic one-degree-of-freedom model, the present approach not only describes well the transient stick-slip dynamics, but it also correctly captures the global stick-slip to stick transition, as well as the secondary bulk motion following the arrest of the contact line, which has been so far overlooked by existing theoretical analyses. This study offers a further contribution to rationalizing the impact of contact angle hysteresis and its associated solidlike friction on the decay of liquid oscillations in the presence of sliding triple lines.
\end{abstract}

\maketitle

\begin{centering}\section{Introduction}\label{sec:C10_Intro}\end{centering}

\bigskip
\begin{centering}\subsection{Linear contact line models for partial wetting conditions}\end{centering}

Liquid sloshing constitutes an archetypal resonator system in fluid mechanics which sometimes represents a critical issue in mechanical engineering and daily life \citep{ibrahim2009liquid,mayer2012walking}. It is therefore crucial to understand the associated damping, as this plays a fundamental role in the mitigation of the maximal wave amplitude response in resonant conditions \citep{bauerlein2021phase,miliaiev_timokha_2023}.\\
\indent Originally the natural frequencies of liquid oscillations in closed basins were derived in the potential flow limit \citep{Lamb32}, while the linear viscous dissipation generated at the free surface, at the solid walls and in the bulk was typically accounted for by a boundary layer approximation \citep{Case1957,Ursell52,Miles67}. This classical theoretical approach is built on the simplifying assumption that the free liquid surface, $\eta$, intersects the lateral wall orthogonally and the contact line can freely slip at a velocity $\partial\eta/\partial t$ with a constant zero slope,
\begin{equation}
\label{eq:FREEcl}
\frac{\partial\eta}{\partial n}=0\ \ \ \  \text{free-end edge condition},
\end{equation}
\noindent where $\partial/\partial n$ is the spatial derivative in the direction normal to the lateral wall. These hypotheses are acceptable for the modelling of gravity-dominated waves in moderately large-size containers, i.e. when capillary effects are negligible \citep{faltinsen2005liquid,bongarzone2022amplitude,marcotte2023super,marcotte2023swirling}, but become questionable when considering smaller-scale containers for which additional dissipations sources originate in the vicinity of the meniscus region, whose dynamics is the central topic of this work.\\ 
\indent With a focus on different contact line conditions, Benjamin \& Scott (1979) \citep{Benjamin79} and Graham-Eagle (1983) \citep{graham1983new} have computed semi-analytically the natural frequencies of liquid oscillations whose contact line is instead fully pinned at the brim of the container,
\begin{equation}
\label{eq:PINNEDcl}
\frac{\partial\eta}{\partial t}=0\ \ \ \ \ \text{pinned-end edge condition},
\end{equation}
\noindent while the interface slope, $\partial\eta/\partial n$, is let free to vary. In this case, theoretical predictions have provided estimations of the system dissipation in better agreement with dedicated experiments \citep{henderson1994surface,martel1998surface,miles1998note,howell2000measurements,nicolas2002viscous,nicolas2005effects,kidambi2009meniscus}. Indeed, with the contact line being fixed, the overall dissipation is ruled by that occurring in the fluid bulk and in the Stokes boundary layers at the bottom and at the solid lateral walls, where the fluid obeys the no-slip condition.\\ 
\indent An intermediate boundary condition that assumes a linear relation between the contact line speed and the slope was proposed by Hocking (1987) \citep{Hocking87},
\begin{equation}
\label{eq:MIXEDcl}
\frac{\partial\eta}{\partial n}=M\frac{\partial\eta}{\partial t}\ \ \ \ \ \text{Hocking condition},
\end{equation}
\noindent with a proportionality constant, sometimes referred to as mobility parameter $M$ \citep{xia2018moving}. According to such a relation, the limiting values $M\rightarrow0$ and $M\rightarrow\infty$ would correspond, respectively, to free-end and pinned-end edge contact line conditions. The agreement with some recent experiments has been found fairly good \citep{li2019stability,bongarzone2023revised}, but the estimation of this proportionality constant is not straightforward \citep{blake1993dynamic,hamraoui2000can,blake2006physics}.\\
\indent The simplicity of these contact line models, which assume that the damping of the system has a linear origin, significantly eases the mathematical tractability of the problem. However, they are too simple to describe the complexity of the region in the neighbourhood of the moving contact line.\\
\indent Improving the modelling of damping effects requires looking more carefully at the dynamics of the oscillating meniscus and at its wetting conditions, a long-standing problem in fluid mechanics that dates back to Navier \cite{navier1823memoire} (see also \cite{Keulegan59,Huh71,Davis1974,miles1990capillary,ting1995boundary,eggers2005existence,Lauga2007,Eral2013} among others).

\bigskip
\begin{centering}\subsection{Nonlinear contact line models for partial wetting conditions}\end{centering}

\begin{figure}[]
    \centering
       \includegraphics[width=0.9\textwidth]{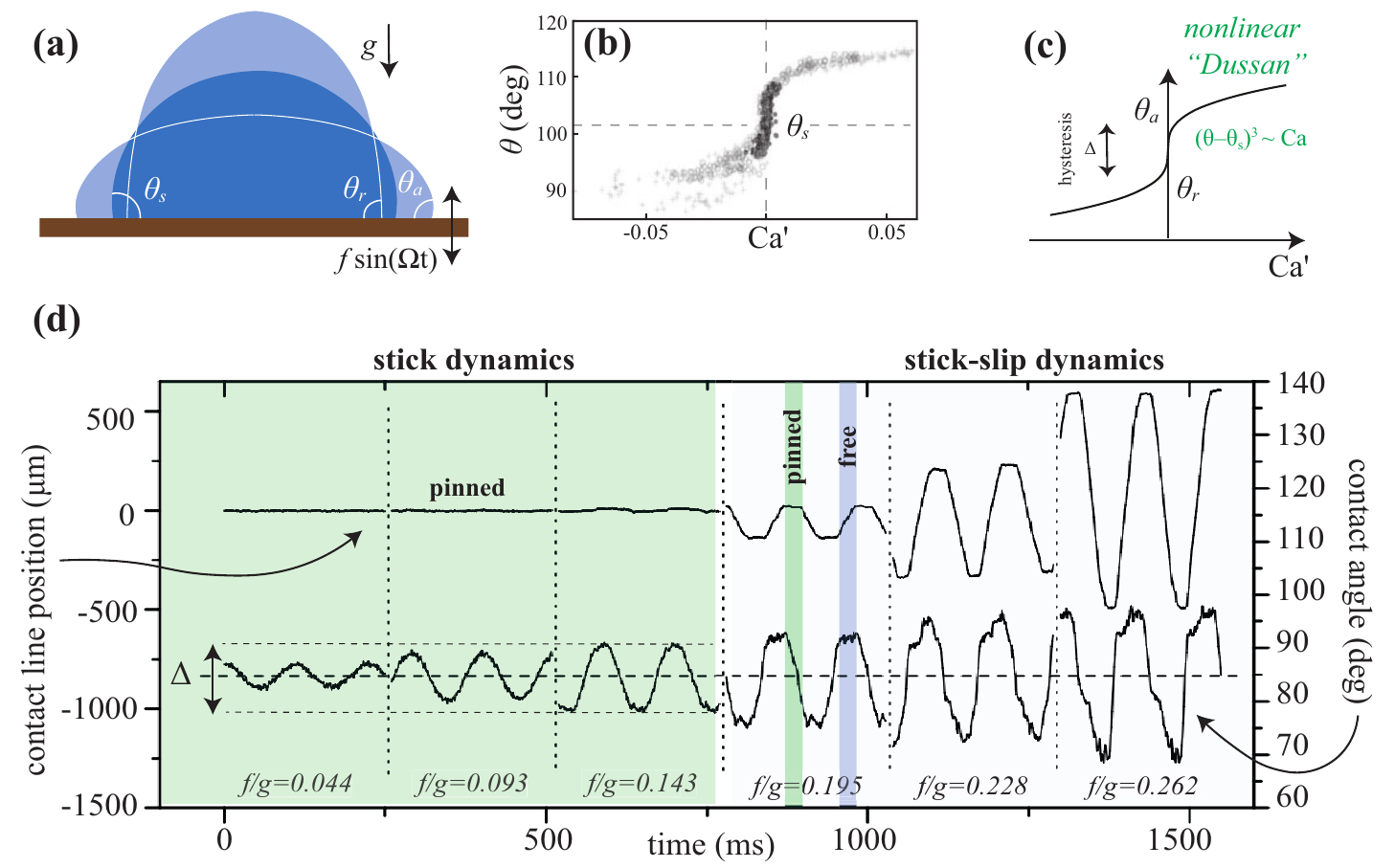}
       \vspace{-0.6cm}
       \caption{(a) Contact angle dynamics in a vertically vibrating droplet. For this oscillatory flows, experiments from (b) Ref.~\onlinecite{xia2018moving} suggest as suitable phenomenological contact angle laws the (c) nonlinear Dussan model \citep{Dussan79,jiang2004contact}. (d) Transition between stick and stick-slip motions in a water sessile drop deposited on a vertically vibrating substrate characterized by a finite contact angle hysteresis ($\Delta\approx10-15$ degrees) \citep{noblin2004vibrated}. Lower curves are contact angle variations versus time, the dashed line represents $\theta_s$. Higher curves are the contact line position around the starting position before vibrations. The six curves for different non-dimensional acceleration amplitudes $f/g$ are joined together in the same plot for comparison. The driving frequency is $1/T= 9\,\text{Hz}$. Panels (b) and (d) are modified versions of figures reported in Refs.~\onlinecite{xia2018moving} and~\onlinecite{noblin2004vibrated}, respectively.}
     \label{fig:P4_1}
\end{figure}

\begin{figure}[!ht!b]
    \centering
       \includegraphics[width=0.9\textwidth]{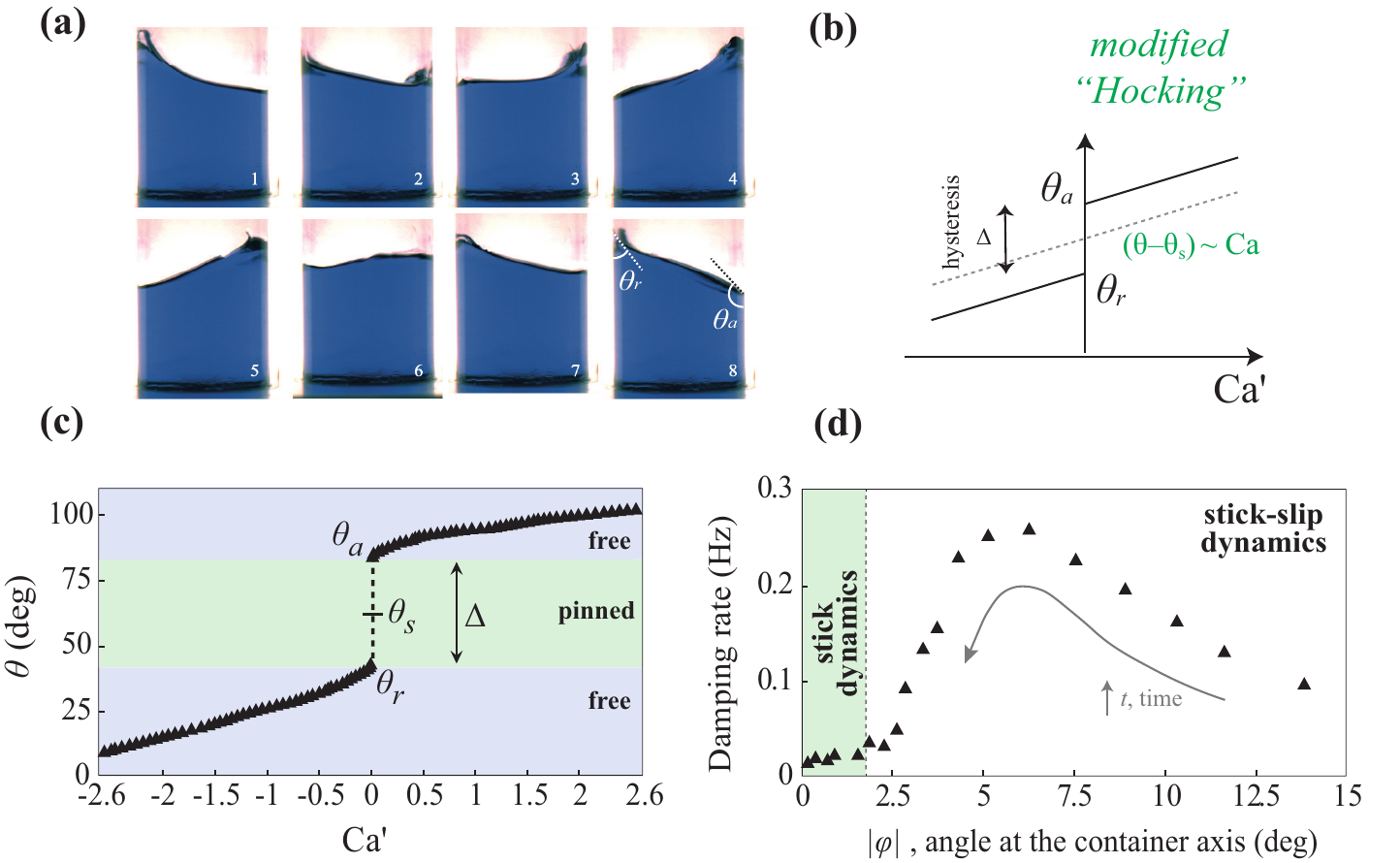}
       \vspace{-0.6cm}
       \caption{(a) Contact angle dynamics in sloshing waves (snapshots over a period) \citep{viola2016resonance}. For this oscillatory flow, experiments from Ref.~\onlinecite{Cocciaro93} suggest as suitable phenomenological contact angle laws the (b) Hocking linear law \citep{Hocking87} supplemented with hysteresis. (c) Experimental contact angle dependence on the capillary number as measured in Ref.~\onlinecite{Cocciaro93} during the natural relaxation dynamics of water oscillations in a cylindrical container initially perturbed. (d) Associated damping rate versus the amplitude of the angle measured at the container axis. The vertical dashed line indicates the value for which the contact line irreversibly pins. Panels (a) and (c)-(d) are modified versions of figures reported in Refs.~\onlinecite{viola2016resonance} and~\onlinecite{Cocciaro93}, respectively.}
     \label{fig:P4_Cocciaro_Viola_model_damp}
\end{figure}

When a liquid meniscus flows over a dry solid substrate, there is a triple-phase interface (air-liquid-solid), which experiences a complex nonlinear dynamics. Experimental observations \citep{Dussan79,rio2005boundary,le2005shape} have shown that the dynamic advancing, $\theta_a$, and receding, $\theta_r$, contact angles deviate from their static values depending on the velocity of displacement of the advancing or receding meniscus. Moreover, there exists a range $\theta\in\left[\theta_r,\theta_a\right]$ within which the contact line seems to remain stationary. The existence of such a static range, defined as contact angle hysteresis, plays a critical role in the nonlinear damping and dynamics of capillary-gravity waves.\\
\indent Several models have been suggested to explain the nonlinear relation between the dynamic contact angles, $\theta$, and the capillary number defined by the contact line velocity, $U$, i.e. $Ca'=\mu U/\gamma$, with $\gamma$ and $\mu$, the air-liquid surface tension and dynamic viscosity, respectively. \citep{voinov1976hydrodynamics,de1985wetting,cox1986dynamics,le2005shape,snoeijer2013moving}.\\
\indent The present investigation focuses on oscillatory flows, for which a brief overview of well-known contact line models is provided in Fig.~\ref{fig:P4_1} and Fig.~\ref{fig:P4_Cocciaro_Viola_model_damp}. For instance, the contact angle dynamics observed for vertical vibrating sessile drops (Fig.~\ref{fig:P4_1}) or during the relaxation of sloshing waves (Fig.~\ref{fig:P4_Cocciaro_Viola_model_damp}) are seen to obey the nonlinear (cubic) \textit{Dussan} model, $\left(\theta-\theta_s\right)^3\sim Ca'$ (see Fig.~\ref{fig:P4_1}(b,c)), and are sometimes well approximated by a modified \textit{Hocking}'s law supplemented with hysteresis (see Fig.~\ref{fig:P4_Cocciaro_Viola_model_damp}(b,c)).
\\
\indent Furthermore, the rich dynamics of an oscillatory meniscus shows some interesting features that the present analysis aims at reproducing and predicting. Some of those features are illustrated in Fig.~\ref{fig:P4_1}(d). In the study conducted by Noblin \textit{et al.} (2004) \citep{noblin2004vibrated}, they investigated the behaviour of a water droplet on a solid surface with a finite contact angle hysteresis under vertical vibration. The results showed two distinct types of oscillations. At low forcing amplitude, the contact line remains pinned and the drop displays eigenmodes at certain resonance frequencies. At higher amplitudes, the contact line starts to move, remaining circular but with a radius oscillating at the excitation frequency. This transition between the two regimes occurs when the variations of the contact angle exceed the hysteresis range. They also observed a decrease in the resonance frequencies at larger vibration amplitudes for which the contact line is mobile. These features were attributed to the hysteresis acting as solidlike friction on the oscillations, leading to a stick-slip regime at intermediate amplitude.\\
\indent In their seminal work, Cocciaro \text{et al.} (1993) \citep{Cocciaro93} thoroughly characterized the contact angle dynamics during the natural (free-of-forcing) relaxation phase of the fundamental asymmetric sloshing mode in a small circular cylindrical container. Two different damping regimes were observed, corresponding to higher and smaller wave amplitude oscillations (see Fig.~\ref{fig:P4_Cocciaro_Viola_model_damp}(d)). First, the contact line slides over the solid substrate experiencing progressive stick-slip transitions under the effect of the dynamic wall friction. In this phase, the damping increases considerably as the wave amplitude decreases, until it reaches a maximum value, after which it starts to decrease, and the small amplitude regime is established. A finite time of arrest for the contact line is found: the interface irreversibly pins and the following pure bulk motion is seen to decay exponentially owing to the linear viscous dissipation acting in the fluid bulk and in the Stokes boundary layers. The natural oscillation frequency initially matches the value associated with a free-end edge eigenmode, it increases during the decay, and it eventually tends to the value associated with a pinned-end edge eigenmode.

\bigskip
\begin{centering}\subsection{Motivation and Objective}\end{centering}

\indent As an alternative to computationally expensive fully nonlinear direct numerical simulations (see \citep{amberg2022detailed,ludwicki2022contact} among others), different theoretical frameworks, attempting to rationalize the nonlinear dependence of the damping rate on the oscillation amplitude, have been recently proposed \citep{Viola2018a,Viola2018b}. These works are based on an asymptotic formulation of the full hydrodynamic problem, which is tackled in the spirit of the weakly nonlinear and multiple timescale approach, under precise assumptions and range of validity. The asymptotic analysis is found to be able to quantitatively predict the nonlinear trend of the damping in the higher amplitudes regime and the existence of a finite-time of arrest for the contact line, in agreement with experiments \citep{Cocciaro93,dollet2020transition}. However, it fails in capturing the transient stick-slip motion and, most importantly, the transition to the small amplitude regime, when the interface pins but the fluid bulk keeps oscillating with a smaller amplitude motion following a purely pinned dynamics.\\
\indent The purpose of the present work is to provide a different theoretical approach, which overcomes the limitations of these asymptotic analyses, thus successfully solving the overall flow dynamics and enabling us to extract and highlight realistic flow features, yet keeping a low computational cost. To this end, we consider liquid oscillations in the simplest sloshing configuration, i.e. liquid columns oscillating in a U-shaped tube, as experimentally investigated by Dollet \textit{et al.} (2020) \citep{dollet2020transition}, and subjected to a physics-inspired nonlinear contact line model following Bongarzone \textit{et al.} (2021) \citep{bongarzone2021relaxation}. Using a piecewise time splitting of the nonlinear contact line law to which the contact line obeys, we formalize a mathematical model based on successive projections between different sets of linear eigenmodes pertaining to each linear split-piece composing the contact line law.\\
\indent The manuscript is organized as follows. In \S\ref{sec:C10_Motiv} we summarize the experimental findings reported by Dollet \textit{et al.} (2020) \citep{dollet2020transition} and comment on the advantages and limitations of the one-degree-of-freedom (1dof) system employed in their study to model the liquid oscillations. We present the full hydrodynamic system in \S\ref{sec:C10_Sec1}, while a numerical characterization in terms of oscillation frequencies and damping rates associated with the various dynamical phases is carried out in \S\ref{sec:C10_Sec2}. The salient points of the projection method presented in Ref.~\onlinecite{bongarzone2021relaxation} are shortly recalled and described in \S\ref{sec:C10_Sec3}. Results and comparison with experiments are given in \S\ref{sec:C10_Sec4}. Lastly, final conclusions are outlined in \S\ref{sec:C10_Conclusion}.

\bigskip
\begin{centering}\section{The case of liquid oscillations in U-shaped tubes}\label{sec:C10_Motiv}\end{centering}

\begin{figure}[]
\centering
\includegraphics[width=1\textwidth]{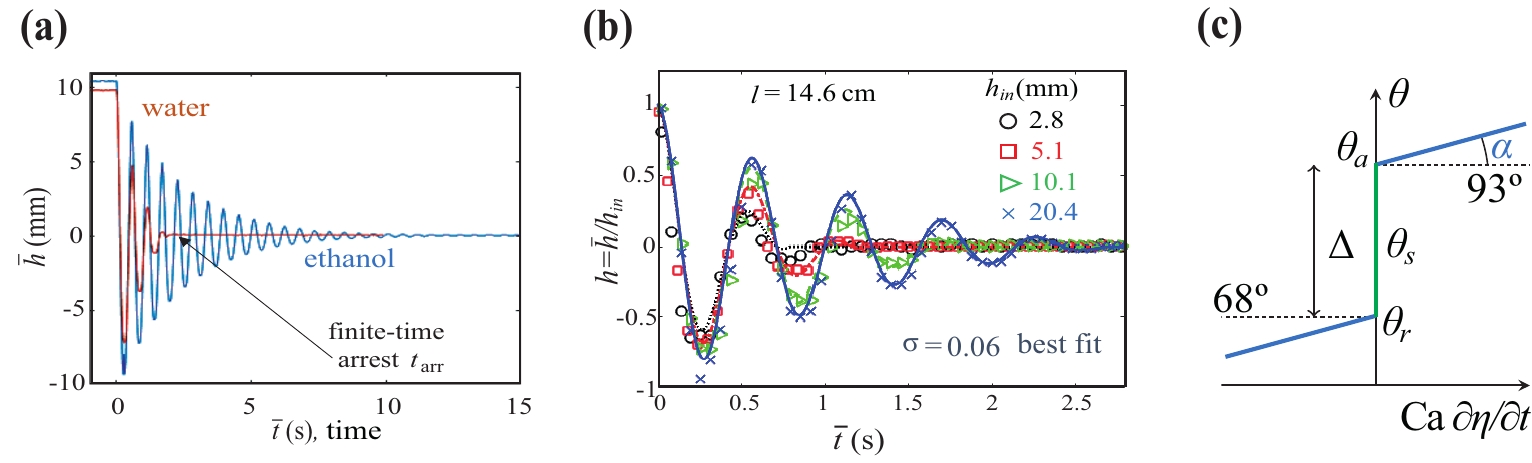}
\caption{(a) Interface height $\bar{h}\left(t\right)\,\text{mm}$ vs time $\bar{t}\left(s\right)$ for water and ethanol in the hydrophobic tube and for liquid column length $l=14.6\,\text{mm}$. (b) Rescaled interface height, $h$, vs time $\bar{t}\left(s\right)$, for water in the hydrophobic tube with a fixed liquid column length $l=14.6\,\text{mm}$ and at different initial elevation $h_{in}$. The solid curves correspond to the predictions from Eq.~\eqref{eq:C10_WNL1dof} with an oscillation period $T=2\pi/\omega_0=2\pi\sqrt{l/2g}$ and with $\sigma=0.06$ as a free fitting parameter common for all experiments. (c) Phenomenological law used in the present work to model the apparent dynamic contact angle, $\theta$, vs the non-dimensional contact line speed, $Ca'=Ca\,\partial\eta/\partial t$, with $Ca=\nu\rho\sqrt{gl/2}/\gamma$, $\nu$ the kinematic liquid viscosity, $\rho$ the liquid density and $\gamma$ the liquid-air surface tension. Panels (a) and (b) are modified versions of figures reported in Ref.~\onlinecite{dollet2020transition}.}
\label{fig:C10_ResDollet}
\end{figure}

\begin{figure}[]
\centering
\includegraphics[width=0.8\textwidth]{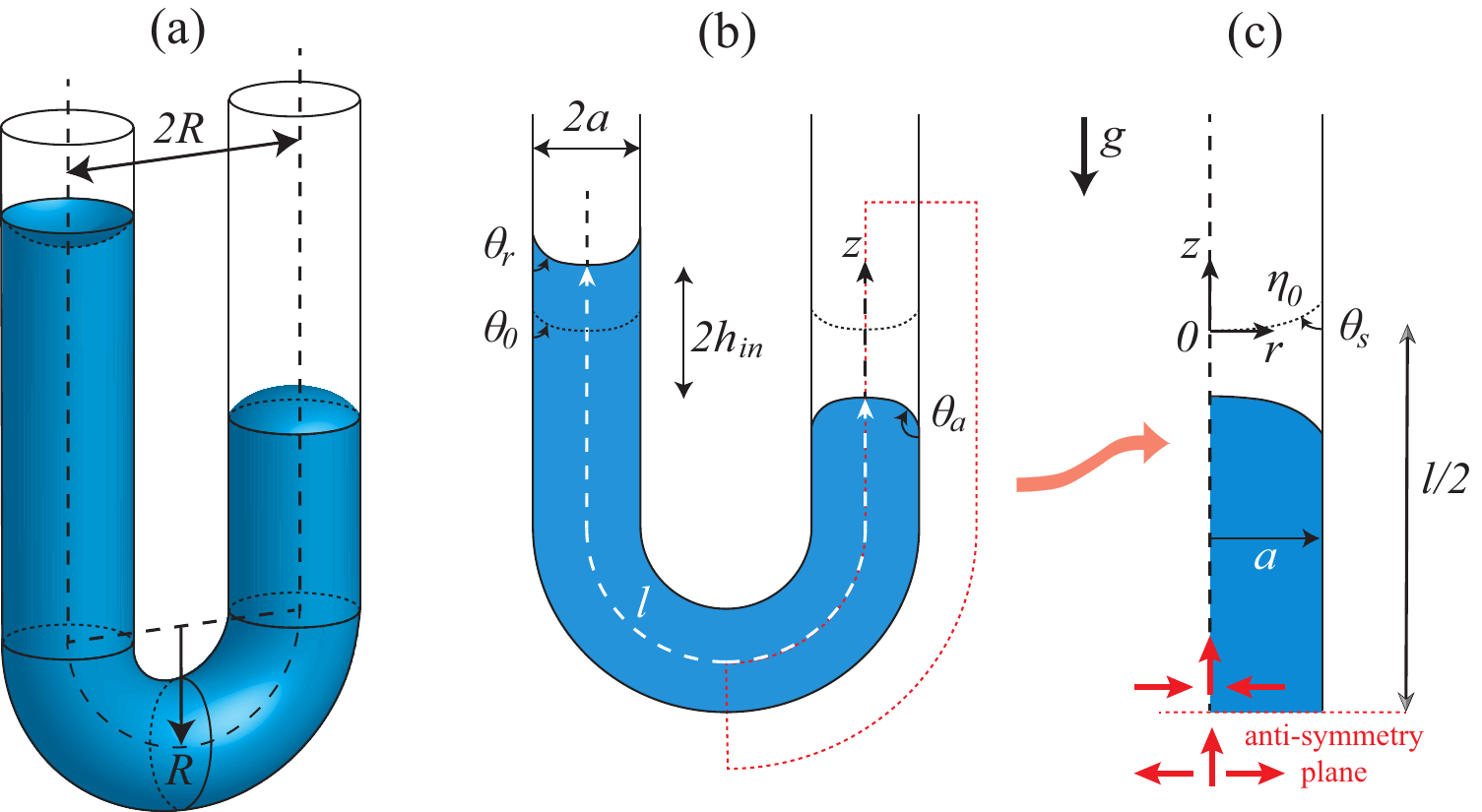}
\caption{Sketch of the U-tube configuration. (a) Full three-dimensional geometry (3D). (b) Two-dimensional (2D) view of the centerline plane. The tube radius is assumed constant and denoted by $a$. The length of the liquid column is $l$. $h$ indicates the height difference of the liquid column between the left and right straight channels. $g$ is the gravity acceleration. The advancing and receding dynamic contact angles are, respectively, $\theta_a$ and $\theta_r$, whereas the static contact angle is labelled as $\theta_s$ and it is in general $\ne90^{\circ}$. (c) If the tube curvature is neglected, the 3D geometry can be reduced to an axisymmetric configuration, by considering only half of the liquid column, of length $l/2$, and by imposing anti-symmetry conditions at the bottom boundary so as to restore the effect of the gravity term on the missing straight channel.}
\label{fig:C10_Sketch}
\end{figure}

Dollet \textit{et al.} (2020) \citep{dollet2020transition} studied the decay of liquid oscillations in a U-shaped tube. They experimentally showed that in the presence of moving contact lines, oscillations are nonlinearly damped, with a finite-time arrest and a dependence on initial conditions. Consistently with the theoretical analysis by Viola \textit{et al.} (2018) \cite{Viola2018a}, they also revealed that contact angle hysteresis can explain this behaviour and quantified the solidlike friction attributable to the contact angle hysteresis.\\
\indent For their experiments, Dollet \textit{et al.} (2020) used two U-shaped glass tubes, one rendered hydrophilic and the other hydrophobic by specifc treatments. The two straight arms of the tubes, separated by a distance $R\approx22.5\,\text{mm}$ (the authors have provided us with this value in a personal communication), have a constant inner radius $a= 8.15\pm0.15\,\text{mm}$ (see Fig.~\ref{fig:C10_Sketch}). Two liquids, namely ultrapure water and absolute ethanol, were used. With regards to the hydrophobic tube, the following wetting properties were measured: $\theta_r=\left(68\pm10\right)^{\circ}$ and $\theta_a=\left(93\pm2\right)^{\circ}$ for water, and $\theta_r=\left(28\pm\right)2^{\circ}$ and $\theta_a=\left(34\pm2\right)^{\circ}$ for ethanol.\\
\indent A controlled volume of liquid, making a column of length $l$ along the centerline, was injected into the tube. Successively, an initial height imbalance $2h_{in}$ between the two contact lines in the left and right straight arms of the tube was introduced and suddenly released. The subsequent natural oscillations of one of the two interfaces were then recorded with a camera.\\
\indent The relaxation of liquid oscillations in the hydrophilic tube, not reported here for the sake of brevity, was observed to be of exponential nature for both ethanol and water. More complex is instead the scenario when dealing with the hydrophobic tube. For this condition, the relevant results of their study are reported in Fig.~\ref{fig:C10_ResDollet}. Panel (a) shows the oscillation decay for both ethanol and water and for the same liquid column length and initial elevation $h_{in}$. For both liquids, the oscillation period, $T$, is well predicted by the analytical formula, i.e. $T=2\pi/\omega_0=2\pi\sqrt{l/2g}$ \citep{Lamb32}, however, for water, the effect of wetting conditions is striking: despite the larger viscosity of ethanol, water oscillations are much more damped, with a finite-time contact line arrest, $t_{arr}$, and a dependence of $t_{arr}$ on the imposed initial condition, $h_{in}$, as illustrated in panel (b).
\\
\indent To rationalize such nonlinear relaxation dynamics for the contact line, the authors employed the 1dof model reminiscent of that of Viola \textit{et al.} (2018) \citep{Viola2018a} and which relies on two assumptions: (i) the tube curvature is neglected and (ii) the flow is hypothesized plug-like. It is difficult to rigorously justify (i), but (ii) appears reasonable as the Stokes boundary layer thickness in these experiments is of the order of $\sqrt{4\pi\nu/T}\approx0.4\,\text{mm}\ll a$($=8.15\,\text{mm}$). This 1dof model then results from the interplay of inertia, gravity as restoring force, linear damping and nonlinear contact line damping included as solid friction:
\begin{subequations}
\begin{equation}
\label{eq:C10_1dof}
\frac{d^2h}{dt^2}+2\sigma \frac{dh}{dt}+h+\mu\,\text{sign}\,\left(\frac{dh}{dt}\right)=0,
\end{equation}
\begin{equation}
\label{eq:C10_mu_alpha}
h=\frac{\bar{h}}{h_{in}},\ \ \ \ t=\omega_0\bar{t},\ \ \ \ \sigma=\frac{\bar{\sigma}\omega_0}{2\pi\rho g a^2},\ \ \ \ \mu=\frac{\gamma\left(\cos{\theta_r-\cos{\theta_a}}\right)}{\rho g a h_{in}},
\end{equation}
\end{subequations}
\noindent with the initial conditions $h=1$ and $dh/dt=0$ at $t=0$ and with the bar symbol denoting dimensional quantities. Importantly, in Eq.~\eqref{eq:C10_1dof}, the linear damping coefficient $\sigma$ is considered as a free-fitting parameter. In the limit of small damping, i.e. $\sigma\ll1$ and $\mu\ll1$, an insightful solution to Eq.~\eqref{eq:C10_1dof} can be obtained by applying the multiple scales method as outlined in Refs.~\onlinecite{Viola2018a,Viola2018b,bongarzone2021relaxation}. The elevation $h\left(t\right)$ is expanded as $h_0+\epsilon h_1\hdots$, with $\epsilon$ a small non-dimensional parameter $\ll 1$ and with a leading order solution $h_0\left(t\right)=\left(1/2\right)A\left(\epsilon t\right)e^{\text{i}t}+\text{c.c.}\,$. Moreover, the amplitude $A\left(\epsilon t\right)$ is assumed to depend on time only through a slow time scale $\sim\epsilon t$. Successively, the imposition of a solvability condition at order $\epsilon$ yields the following asymptotic approximation,
\begin{equation}
\label{eq:C10_WNL1dof}
h\left(t\right)=\left[-\frac{2\mu}{\pi\sigma }+\left(1+\frac{2\mu}{\pi\sigma}\right)e^{-\alpha t/2}\right]\cos{t},
\end{equation}
if $t\le t_{arr}$, and $h=0$ if $t\ge t_{arr}$, with $t_{arr}=\frac{1}{\sigma}\log{\left[1+\left(\pi\sigma/2\mu\right)\right]}$ the time of arrest of the contact line oscillations. Eq.~\ref{eq:C10_WNL1dof} predicts an envelope shape that varies from the classical exponential damping as $\sigma\gg\mu$ (nearly linear dissipation) to a linear decay in time as $\mu\gg\sigma$ (solidlike friction). In spite of the strong oversimplifications, the 1dof model predicts fairly well the experimental contact line dynamics once the damping $\sigma$ is fitted from experiments. In the experimental range of liquid column lengths explored, a unique value of $\sigma$, i.e. $\sigma=0.06$ (for water), allowed for a good overall comparison. One can therefore state that the 1dof nonlinear pendulum-like model is capable of reproducing the global features of the relaxation dynamics in the presence of contact angle hysteresis, hence providing a powerful tool to obtain a quick estimation, e.g., of the finite-time arrest.\\
\indent Nevertheless, a few main limitations are worth to be commented on. Preceding the time of arrest, the contact line exhibits some transient stick-slip transitions (visible in Fig.~\ref{fig:C10_ResDollet}(a) and~(b)). As discussed in Ref.~\onlinecite{bongarzone2021relaxation}, each time that the contact line transiently reaches a zero speed, the contact angle will have to adjust from $\theta_a$ to $\theta_r$ (or vice versa) while the contact line remains pinned; this dynamical variation obviously requires a certain time-interval to happen. Most importantly, after the time of arrest, the fluid bulk still exhibits oscillations, even if the contact line is pinned. These secondary oscillations are unaffected by nonlinear friction and, therefore, decay exponentially under the effect of pure linear viscous dissipation (see Supplementary Material of Ref.~\onlinecite{dollet2020transition} for an experimental quantification of the damping rate and frequency in the pinned regime). Such a stick-slip-to-stick transition cannot be captured by a the 1dof model, as it intrinsically calls for a modelization of the many-degrees-of-freedom of the system. Lastly, the 1dof model requires the fitting of the linear damping, $\sigma$, whose accurate computation can be very subtle. The linear damping englobes multiple dissipative effects: the dissipation occurring in the Stokes boundary laters at the tube walls, the one induced by three-dimensional effects in the curved part of the tube and, particularly, possible extra dissipation sources linked to the contact line motion, such as a dynamical contact angle variation at a non-zero contact line speed (see Fig.~\ref{fig:C10_ResDollet}(c)) which is a ubiquitous feature of similar experiments (see, for instance, Refs.~\onlinecite{Hocking87,Cocciaro93,jiang2004contact,rio2005boundary,snoeijer2013moving,xia2018moving,fiorini2022effect}, among others).\\
\indent With the aim of building a more refined model so as to overcome these limitations, in the following we will characterize the present U-tube dynamics by considering the full hydrodynamic system of governing equations, to which we will apply the projection method developed by Bongarzone \textit{et al.} (2021) \citep{bongarzone2021relaxation}. The case of water oscillations in the hydrophobic tube described in Dollet \textit{et al.} (2020) and summarized in Fig.~\ref{fig:C10_ResDollet} will represent our experimental reference condition.

\bigskip
\begin{centering}\section{Full Hydrodynamic System}\label{sec:C10_Sec1}\end{centering}

\begin{centering}\subsection{Governing equations}\end{centering}

With regards to the experimental setup of Ref.~\onlinecite{dollet2020transition} previously discussed, let us consider a U-shaped tube of radius $a$ and filled with a liquid column of length $l$, as illustrated in Fig.~\ref{fig:C10_Sketch}(a,b). The section of the tube is assumed constant all over the tube length, a first geometrical approximation already dealt with by Dollet \textit{et al.} (2020) \citep{dollet2020transition}. The geometry of the problem remains intrinsically three-dimensional (3D). Nevertheless, by analogy with the approach employed by Iguchi \textit{et al.} (1982) \citep{iguchi1982analysis} and Dollet \textit{et al.} (2020), in the following, we neglect the tube curvature. This is certainly a strong \textit{a priori} assumption, which appears worth to be discussed. Appendix~\ref{sec:C10_AppA} is devoted to discussing, at least partially, its justification. Under this hypothesis, one may then imagine cutting the tube in half and unfolding it, so as to consider the $z$-axis as straight and only half of the liquid column, of length $l/2$. At this stage, we have reduced the 3D geometry to an axisymmetric configuration, that can now be more easily described in cylindrical coordinates, $\text{O}r\phi z$. The origin of the cylindrical reference system is located at the intersection of the unperturbed static interface at $z=\eta_0$ with the centerline axis at $r=0$. The effect of the gravity term on the missing half of the domain can be correctly restored by considering proper anti-symmetry conditions on the bottom boundary at $z=-l/2$ (Fig.~\ref{fig:C10_Sketch}(c)). The sudden sign switching of the effect of gravity in $z=-l/2$ is consistent with neglecting the curvature in the U-turn region.\\
\indent The viscous flow within the U-shape tube is thus governed by the incompressible Navier-Stokes equations 
\begin{equation}
\label{eq:C10_GovEq}
\nabla\cdot\mathbf{u}=0,\ \ \ \frac{\partial \mathbf{u}}{\partial t}+\left(\mathbf{u}\cdot\nabla\right)\mathbf{u}+\nabla p-\frac{1}{Re}\Delta\mathbf{u}=-1\hat{\mathbf{e}}_z,
\end{equation}
which are made nondimensional by using the container's characteristic length $l$ and the velocity $\sqrt{gl/2}$ (Fig.~\ref{fig:C10_Sketch}). Consequently, the Reynolds number is defined as $Re=\frac{\sqrt{g \left(l/2\right)^3}}{\nu}$ and the term $-1\hat{\mathbf{e}}_z$ denotes the nondimensional gravity acceleration. In Eq.~\eqref{eq:C10_GovEq}, $p\left(r,z,t\right)$ is the pressure field, whereas $\mathbf{u}\left(r,z,t\right)=\left\{u,w\right\}^T$ is the velocity field, with $u$ and $w$ the radial and axial velocity, respectively. Note that the dynamics is assumed axisymmetric and such assumption will be maintained throughout the manuscript. At the free surface, $z=\eta$, kinematic and dynamic boundary conditions hold,
\begin{subequations}
\begin{equation}
\label{eq:C10_KIN}
\frac{D\left(\eta-z\right)}{Dt}=\frac{\partial\eta}{\partial t}+u\,\frac{\partial\eta}{\partial r}-w=0,
\end{equation}
\begin{equation}
\label{eq:C10_DYN}
\left[-p\mathbf{I}+\frac{1}{Re}\left(\nabla\mathbf{u}+\nabla^T\mathbf{u}\right)-\frac{1}{Bo}\kappa\left(\eta\right)\mathbf{I}\right]\cdot\mathbf{n}=\mathbf{0},
\end{equation}
\end{subequations}
where $D/Dt$ is the material derivative, $\mathbf{n}=\left(1+\eta_r^{2}\right)^{-1/2}\left\{-\eta_r,1\right\}^T$ is unit vector normal to the interface, and $\kappa$ is the free surface curvature, $\kappa\left(\eta\right)=\left[\eta_{rr}+r^{-1}\eta_r\left(1+\eta_r^2\right)\right]\,\left(1+\eta_r^2\right)^{-3/2}$. The Bond number is defined as $Bo=\frac{\rho ga^2}{\gamma}\left(\frac{l/2}{a}\right)^2$, with $\gamma$ designating the air-liquid surface tension. As anticipated above, the restoring effect of the missing half of the tube is reintroduced by imposing anti-symmetry conditions for $u$ and $w$ at the bottom boundary (see Fig.~\ref{fig:C10_Sketch}(c)). More precisely, we impose
\begin{equation}
u=\frac{\partial w}{\partial z}=0 \ \ \ \ \text{at $z=-1$}.
\end{equation}
\noindent Moreover, owing to the axisymmetric assumption, the axis boundary condition imposes
\begin{equation}
\label{eq:C10_axisBC}
u=\frac{\partial w}{\partial r}=0\ \ \ \ \text{at $r=0$}.
\end{equation}

\bigskip
\begin{centering}\subsection{Treatment of the sidewall: a macroscopic depth-dependent slip-length model}\end{centering}

\textcolor{black}{With regards to the modelling of the sidewall boundary condition, the case of a pinned contact line is compatible with the classical no-slip condition \citep{bongarzone2022sub}. The latter will be employed throughout the paper whenever dealing with a fixed contact line. On the other hand, the no-slip condition and a moving contact line are not compatible with each other and one must adopt different strategies.}\\
\indent Here we adopt a slip-length model, thus assuming that the fluid speed relative to the solid wall is proportional to the viscous stress \citep{navier1823memoire,Lauga2007} and that, together with the no-penetration condition, provides the boundary conditions
\begin{equation}
\label{eq:C10_Slip_condition}
u=0,\ \ \ w+ l_s\left(z\right)\frac{\partial w}{\partial x}=0\ \ \ \text{at $r=\frac{a}{l/2}$}.
\end{equation}
Such a condition is indeed needed in order to regularize the stress singularity at the moving contact line \citep{Huh71,Davis1974}. It was hypothesized by Miles (1990) \citep{miles1990capillary} and Ting \& Perlin (1995) \citep{ting1995boundary} that the phenomenological macroscopic slip length appearing in Eq.~\eqref{eq:C10_Slip_condition} should not be assumed constant along the wall, but rather spatially dependent on the position along the lateral wall and vanishing at a certain distance away from the contact line, where the flow obeys the no-slip condition. For this reason, we employ here a depth-dependent slip length as proposed by Bongarzone \& Gallaire (2022) \citep{bongarzone2022numerical}, which has been shown to correctly estimate the linear dissipation occurring in the Stokes boundary layers at the lateral solid walls (see Appendix~\ref{sec:C10_AppB} for further validations specific to the present case). Briefly, we postulate that the slip length $l_s\left(z\right)$ is described by the exponential law
\begin{equation}
\label{eq:C10_AppB_lz2}
l_s\left(z\right)=l_{cl}\,\text{exp}\left(-\frac{z}{\delta}\text{log}\left(\frac{l_{\delta}}{l_{cl}}\right)\right),\ \ \ \ z\in\left[-H,0\right].
\end{equation}
\noindent In Eq.~\eqref{eq:C10_AppB_lz2}, $l_{cl}$ is the slip-length value at the contact line, $r=a/\left(l/2\right)$ and $z=0$, whereas $l_{\delta}$ is its value at a distance $\delta$ below the contact line, $r=a/\left(l/2\right)$ and $z=-\delta$, with $\delta$ representing the size of the slip region \citep{ting1995boundary}. In principle, $l_{cl}$, $l_{\delta}$ and $\delta$ are all free parameters. However, keeping in mind that, macroscopically speaking, one aims at mimicking a stress-free condition in the vicinity of the contact line and a no-slip condition after a certain distance $\delta$, the natural choice is $l_{cl}\gg1$ ($\sim 10^{2}$$\div$$10^{4}$) and $l_{\delta}\ll1$ ($\sim 10^{-4}$$\div$$10^{-6}$). The range of values proposed in brackets is based on the sensitivity analysis reported in Ref.~\onlinecite{bongarzone2022numerical}, whereas the slip region penetration depth, $\delta$, as postulated by Miles (1990) \citep{miles1990capillary}, is here assumed of the order of the non-dimensional Stokes boundary layer thickness, i.e. $\delta\approx \left(l/2\right)^{-1}\delta_{St}=\left(l/2\right)^{-1}\sqrt{2\nu/\omega_0}$, with $\omega_0^2=2g/l$. \textcolor{black}{What mostly matters is that $\delta$ is kept small with respect to all other scales at hand in the problem, i.e. $l$, $a$, $R$, capillary length $\sqrt{\gamma/\rho g}$ or Stokes boundary layer thickness $\sqrt{2\nu/\omega_0}$.}

\bigskip
\begin{centering}\subsection{Phenomenological contact angle model and static meniscus}\end{centering}

Lastly, to model the contact line motion, $z=\eta$ and $r=a/\left(l/2\right)$, we include the phenomenological law of Fig.~\ref{fig:C10_ResDollet}(f), which describes the nonlinear contact angle dynamic as a function of the contact line speed,
\begin{equation}
\label{eq:C10_geom_rel_cl}
\frac{\partial\eta}{\partial r}=\pm\cot{\theta},\ \ \ \theta-\theta_s=\alpha Ca\frac{\partial\eta}{\partial t}+\frac{\Delta}{2}\ \text{sign}\left(\frac{\partial\eta}{\partial t}\right)\ \ \ \ \ \left(\text{Hocking+hysteresis}\right),
\end{equation}
\noindent with $Ca=\nu\rho\sqrt{gl/2}/\gamma$ and with the value of $\alpha$ that will be discussed and specified in the next section. Note that this model has already been used in Ref.~\onlinecite{bongarzone2021relaxation} and it results from a combination of the linear Hocking's law \citep{Hocking87}, of slope $\alpha$, and a static contact angle hysteresis of range $\Delta$. In the rest of the paper, to simplify calculations, we will additionally (and somewhat naively) assume that the advancing and receding phases are completely symmetric and that the hysteresis range is centered around $\theta_s$, i.e. $\theta^+=\theta_a-\theta_s=\Delta/2$ and $\theta^-=\theta_r-\theta_s=-\Delta/2$, while being aware that the advancing and receding contact line dynamics are generally characterized by different value of $\alpha$, i.e. $\alpha_A\ne\alpha_R$ \citep{voinov1976hydrodynamics,Dussan79,de1985wetting,cox1986dynamics,rio2005boundary,Cocciaro93}.
\begin{figure}[]
\centering
\includegraphics[width=0.75\textwidth]{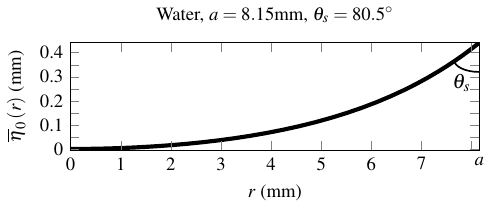}
\caption{Shape of the dimensional static meniscus, $\overline{\eta}_0$, computed numerically for $\theta_s=\left(\theta_a+\theta_r\right)/2=\left(93+68\right)/2=80.5^{\circ}$.}
\label{fig:C10_Fig2}
\end{figure}\\
\indent In the limit of small oscillation amplitudes and small static contact angle hysteresis, the fully nonlinear governing equations~\eqref{eq:C10_GovEq} together with their boundary conditions~\eqref{eq:C10_KIN}-\eqref{eq:C10_geom_rel_cl} can be linearized around the rest state, characterized by zero velocity and pure hydrostatic pressure. With regards to the experiments by Dollet \textit{et al.} (2020) for water in the hydrophobic tube, the measured advancing and receding contact angles are, respectively, $\theta_a=93^{\circ}$ and $\theta_r=68^{\circ}$. If we hypothesize the equilibrium angle $\theta_s$ to be the averaged value of $\theta_a$ and $\theta_r$, this amounts to $\theta_s=80.5^{\circ}$, meaning that the static free surface is not flat (as it would be for $\theta_s=90^{\circ}$). We therefore linearize the system of equations around an initially curved static meniscus, whose resulting axisymmetric shape, reported in Fig.~\ref{fig:C10_Fig2}, is computed as the solution of the following static equation:
\begin{equation}
\label{eq:C10_StaticMen}
\eta_0=\frac{1}{Bo}\left[\frac{\eta_{0,rr}+r^{-1}\eta_{0,r}\left(1+\eta_{0,r}^2\right)}{\left(1+\eta_{0,r}^2\right)^{3/2}}\right],\ \ \ \text{with}\ \ \ \left.\frac{\partial\eta_0}{\partial r}\right|_{r=0}=0, \ \ \ \left.\frac{\partial\eta_0}{\partial r}\right|_{r=a/\left(l/2\right)}=\cot{\theta_s},
\end{equation}
\noindent Eq.~\eqref{eq:C10_StaticMen} is nonlinear in $\eta_0$ and can be solved numerically using an iterative Newton method as described in Appendix~A.1 of Ref.~\onlinecite{Viola2018a}.

\bigskip
\begin{centering}\section{Natural properties of the system}\label{sec:C10_Sec2}\end{centering}

Notwithstanding the linearization of the governing equations around the rest state, the system is still nonlinear owing to the hysteretic contact angle model~\eqref{eq:C10_geom_rel_cl}. Nevertheless, it appears intuitive that the underlying contact line motion can be split into two distinct dynamical phases, namely a pinned-phase, described by the condition
\begin{equation}
\label{eq:C10_pinn_bc}
\frac{\partial \eta}{\partial t}=0\ \ \ \text{(pinned-phase)},
\end{equation}
\noindent and a free-phase with 
\begin{equation}
\label{eq:C10_free_bc}
\frac{\partial \eta}{\partial r}+\alpha Ca\frac{\partial\eta}{\partial t}=-\theta^{\pm}\ \ \ \text{(free-phase)},
\end{equation}
\noindent both evaluated at $r=a/\left(l/2\right)$. The non-homogeneous term in the right-hand side of Eq.~\eqref{eq:C10_free_bc} will be dealt with within the formalism of the projection method. Let us ignore this term for the moment by rewriting
\begin{equation}
\label{eq:C10_free_bc_unf}
\frac{\partial \eta}{\partial r}+\alpha Ca\frac{\partial\eta}{\partial t}=0.
\end{equation}
\noindent Then, the system of governing equations closed by these two boundary conditions, taken independently, translate into two separated fully linear homogeneous problems, that can be both written in the form 
\begin{equation}
\label{eq:C10_lin_prob}
\mathcal{B}_{f,p}\frac{\partial}{\partial t}\mathbf{q}_{f,p}=\mathcal{A}_{f,p}\mathbf{q}_{f,p}.
\end{equation}
\noindent with $\mathbf{q}_{f,p}=\left\{\mathbf{u}_{f,p},p_{f,p},\eta_{f,p}\right\}^T$ the state vector. The symbolic expressions of the mass matrix $\mathcal{B}_{f,p}$ and the stiffness matrix $\mathcal{A}_{f,p}$ are explicitly given in Ref.~\onlinecite{bongarzone2021relaxation}, while the subscripts $_{f,p}$ are here used to designate either the free ($_f$) or the pinned ($_p$) phase. By introducing the ansatz
\begin{equation}
\label{eq:C10_ansatz_eig}
\mathbf{q}_{f,p}=\hat{\mathbf{q}}_{f,p}e^{\lambda_{f,p}t}+c.c.\,,
\end{equation}
\noindent with $\lambda_{f,p}=-\sigma_{f,p}+\text{i}\omega_{f,p}$, equation~\eqref{eq:C10_lin_prob} reduces to the following generalized eigenvalue problem
\begin{equation}
\label{eq:C10_gen_eig}
\lambda_{f,p}\mathcal{B}_{f,p}\hat{\mathbf{q}}_{f,p}=\mathcal{A}_{f,p}.
\end{equation}
\noindent Matrices $\mathcal{A}_{f,p}$ and $\mathcal{B}_{f,p}$ are numerically discretized by means of a Chebyshev collocation method implemented in Matlab in the same fashion of Refs.~\onlinecite{Viola2018a,Viola2018b,bongarzone2021relaxation,bongarzone2022numerical}; the resulting eigenvalue problem is also solved in Matlab via the built-in \textit{eigs} function.
\begin{figure}[]
\centering
\includegraphics[width=0.8\textwidth]{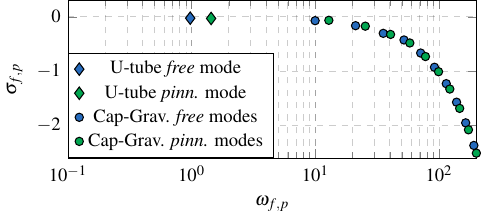}
\caption{Eigenvalue spectrum associated with the two contact line boundary conditions, i.e. pinned (green markers) and free (blue markers), computed numerically by solving the generalized eigenvalue problem~\eqref{eq:C10_gen_eig}. For the case of a free contact line condition, the calculation here reported has been performed by imposing a value of $\alpha=0$. Both spectra are computed for a liquid column length $l=14.6\,\text{cm}$. Fluid properties: water, $\rho=1000\,\text{kg/m$^3$}$, $\gamma=0.0725\,\text{N/m}$ and $\nu=1\times10^{-6}\,\text{m$^2$/s}$.}
\label{fig:C10_FIg19}
\end{figure}
\begin{figure}[]
\centering
\includegraphics[width=1\textwidth]{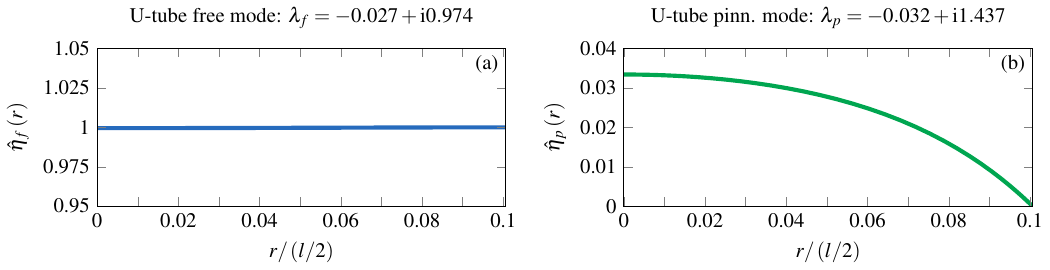}
\caption{(a) Eigen-interface associated with the U-tube free mode computed in~\ref{fig:C10_FIg19}. The free surface dynamics in the free-phase consists of an upward-downward oscillation of a flat interface. (b) Eigen-interface associated with U-tube pinned mode computed in~\ref{fig:C10_FIg19}. Instead, the surface dynamics in the pinned-phase consists of an interface oscillating with a bell-like shape whose edges are anchored at the wall.}
\label{fig:C10_FIg17bis}
\end{figure}\\
\indent The eigenvalue spectrum associated with the solution of the two independent eigenvalue problems is reported in Fig.~\ref{fig:C10_FIg19}. This figure shows, for both wetting phases, a spectrum that contains two families of oscillating natural modes, namely a free/pinned U-tube mode and free/pinned capillary-gravity waves. However, these waves oscillate at a much larger frequency, at least ten times higher, than the fundamental U-tube mode, and are typically more damped than the U-tube mode. The latter mode, with its dynamical properties and structure, displayed in Fig.~\ref{fig:C10_FIg17bis}, is, therefore, the mode that is expected to govern the dynamics.\\
\indent Hence, in the next two sub-sections we will carefully comment on the eigenvalue properties of such U-tube modes, tackled separately in the two dynamical phases. For simplicity, we will start from the pinned-phase, which appears easily describable from a numerical perspective. Successively, we will handle the free-phase, whose description hinges on the subtle modelling of the moving contact line and slip length conditions.

\bigskip
\begin{centering}\subsection{Pinned-phase}\label{subsec:C10_Sec2sub1}\end{centering}

\begin{figure}[]
\centering
\includegraphics[width=0.65\textwidth]{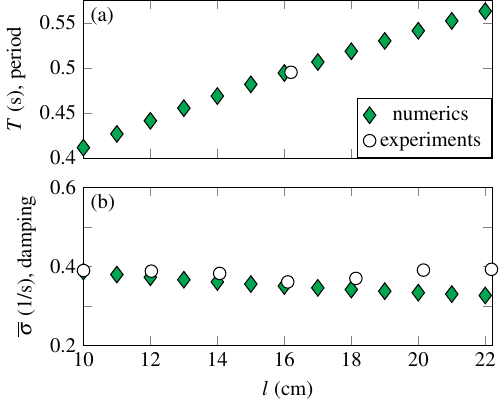}
\caption{Dimensional oscillation period, $T$, and damping coefficient, $\overline{\sigma}$, versus the water column length and associated with a pinned contact line dynamics of the fundamental U-tube mode. Green diamonds: values computed fully numerical eigenvalue calculation. White circles: values measured experimentally as reported in Ref.~\onlinecite{dollet2020transition}.}
\label{fig:C10_Fig10}
\end{figure}

The dependence of the oscillation period and of the damping coefficient on the liquid column length for the U-tube pinned mode, as numerically computed, is shown in Fig.~\ref{fig:C10_Fig10}. Only one experimental value has been reported by Dollet \textit{et al.} (2020) \citep{dollet2020transition} (in their Supplementary Material) and it seems in agreement with our trend, which is also reminiscent of that displayed in Fig.~\ref{fig:C10_ResDollet}(c), although no analytical dispersion relation exists for a pinned contact line.\\
\indent More experimental values are available with regard to the damping coefficient. Although some discrepancies are observed at larger values of $l$, an overall fair agreement is found when compared with our numerical estimates.\\
\indent In this regard, it is important to realize that a pinned contact line condition is mathematically fully compatible with a no-slip wall condition, i.e. no stress singularity needs to be resolved at the contact line, hence allowing one for a precise numerical estimation of the damping. If we ignore experimental errors and ensure numerical convergence, the main possible source of disagreement with these experiments is attributable to free surface contamination or three-dimensional (3D) effects, overlooked by our ideal axisymmetric model, which neglects the tube curvature. To be sure that 3D effects are not important, in Appendix~\ref{sec:C10_AppA}, we perform a full 3D eigenvalue calculation so as to refine the numerical values reported in Fig.~\ref{fig:C10_Fig10}. This calculation proves 3D corrections to be small.

\bigskip
\begin{centering}\subsection{Free-phase}\label{subsec:C10_Sec2sub2}\end{centering}

\begin{centering}\subsubsection{Ignoring dynamical contact angle variation: $\alpha=0$}\end{centering}

\begin{figure}[]
\centering
\includegraphics[width=0.65\textwidth]{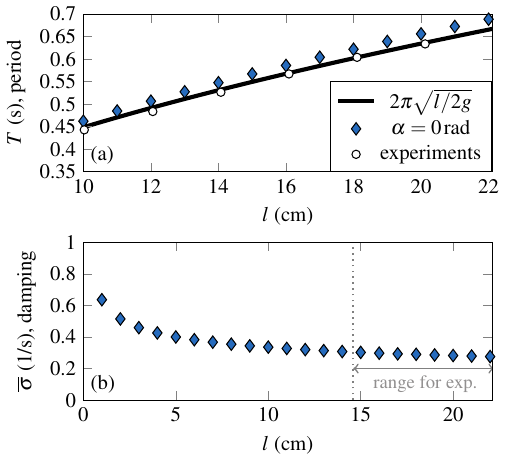}
\caption{(a) Dimensional oscillation period, $T$, and (b) damping coefficient, $\overline{\sigma}$, versus the water column length, $l\left(cm\right)$ and associated with a free contact line dynamics of the fundamental U-tube mode. Blue diamonds: values computed fully numerical eigenvalue calculation by accounting for the variable slip length model discussed in Eq.~\eqref{eq:C10_free_bc_unf} with $\alpha=0$. White circles in (a): values measured experimentally as reported in Ref.~\onlinecite{dollet2020transition}. The experimental range investigated in Ref.~\onlinecite{dollet2020transition} is indicated by the grey arrow in (b). Within this range, the damping coefficient is nearly constant with the tube length.}
\label{fig:C10_FIg13}
\end{figure}

By analogy with the pinned case, the dependence of the oscillation period and of the damping coefficient on the liquid column length for the U-tube free mode is shown in Fig.~\ref{fig:C10_FIg13}. The numerics slightly overestimate the oscillation period, but overall it is in good agreement with the experiments. The fact that the experimental data are better described by the theoretical formula, which does not account for viscous dissipation, is however counter-intuitive. Pure viscous dissipation should indeed introduce a viscous correction to the natural frequency, which should result in a diminished value or, equivalently, in a higher oscillation period $T$. This may suggest that there is a second effect counteracting and compensating for such a viscous correction to the natural frequency. Appendix~\ref{sec:C10_AppA} shows that, among the small three-dimensional effects ignored in the present analysis, the curved part of the U-tube may lead to a small increase in the natural frequencies that can contribute to this compensation effect.\\
\indent In employing the 1dof model, Dollet \textit{et al.} (2020) used a non-dimensional linear damping coefficient $\sigma$ fitted from experiments and whose best-fit value amounts to $0.06$. This coefficient is difficult to predict precisely, as it englobes several contributions, among which is the dissipation occurring in the laminar Stokes boundary layers at the lateral walls.\\
\indent The numerical approach here employed, based on the slip length model previously discussed, provides a tool to compute the dissipation associated with the Stokes boundary layers (see Ref.~\onlinecite{bongarzone2022numerical} for further details).\\
\indent Fig.~\ref{fig:C10_FIg13}(b) shows that within the experimental range of liquid column length, $l\left(cm\right)$, considered, the damping $\sigma$ does not vary much with $l$, thus possibly explaining why a single value of $\sigma$ fitted from experiments can allow a good match with those measurements. The present numerical calculation for the damping is also compared to an analytical estimate developed in Appendix~\ref{sec:C10_AppB}, that also validates the numerical scheme.\\
\indent Nevertheless, the non-dimensional averaged value in the experimental range of water column lengths, amounts to $\sigma\approx0.027$, which is less than half the one needed for a good agreement with the data. The averaged value is computed as $\sigma=n_i^{-1}\sum_{i}^{n_i}\overline{\sigma}_i\sqrt{l_i/2g}$, with $n_i$ the number of lengths $l$ used to sample the experimental range.\\
\indent As discussed in Appendix~\ref{sec:C10_AppA}, three-dimensional effects related to the tube curvature can produce an increase in the damping of a few percentages, but this is not sufficient to explain such a mismatch. The extra dissipation missing in the modelization of the free phase is therefore very likely attributable to the contact line dynamics.

\bigskip
\begin{centering}\subsubsection{Accounting for dynamical contact angle variation: $\alpha\ne0$}\end{centering}

\begin{figure}[]
\centering
\includegraphics[width=0.65\textwidth]{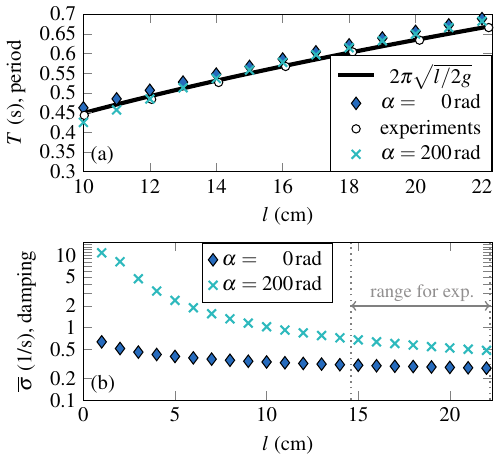}
\caption{Same as in Fig.~\ref{fig:C10_FIg13} (here in $\sigma$-$\log$ scale), but with the light blue crosses indicating the values computed by also accounting for extra contact line dissipation produced by Hocking's law \citep{Hocking87,bongarzone2021relaxation} with $\alpha=200\,\text{rad}$). Within this range, the damping coefficient is nearly constant with the tube length, $l$, even for $\alpha=200\,\text{rad}$. The average value in this range is $\sigma\approx0.06$, which matches the one used in Fig.~\ref{fig:C10_ResDollet} and obtained from the best-fit of the experiments.}
\label{fig:C10_FIg13bis}
\end{figure}

\textcolor{black}{As in the experimental conditions considered here the extra contact line dissipation is well englobed into a linear damping coefficient, we propose to adopt a linear law for the dynamic contact angle variations being proportional to the contact line speed.} We therefore reintroduce the contact line parameter that characterizes the Hocking law, i.e. $\alpha\ne0$. Recalling the contact line condition for the free-phase~\eqref{eq:C10_free_bc_unf}, one can see how a value of $\alpha=0$ would correspond to a contact line sliding over the solid substrate with a constant and zero slope (dashed lines in Fig.~\ref{fig:C10_ResDollet}). On the other hand, the pinned condition~\eqref{eq:C10_pinn_bc} is nothing more than a limiting case of Eq.~\eqref{eq:C10_free_bc} with $\alpha\rightarrow+\infty$. We are supposing here to be in an intermediate situation where $\alpha$, sometimes also referred to as friction coefficient \citep{hamraoui2000can} or mobility parameter $M$ \citep{xia2018moving}, assumes a finite value different from zero.\\ 
\indent Let us first blindly consider $\alpha$ as a free fitting parameter. A value of $\alpha=200\,\text{rad}$ leads to a non-dimensional averaged (in the experimental range of Fig.~\ref{fig:C10_FIg13bis}) damping coefficient of $\sigma=\overline{\sigma}\sqrt{l/2g}\approx0.06$, which is exactly the value that was fitted by Dollet \textit{et al.} (2020). If this procedure shows that a simple linear dynamic contact line model is sufficient to explain the missing dissipation, one can wonder whether the value of $\alpha$ used is meaningful for the experimental conditions discussed here.\\
\indent Hamraoui \textit{et al.} (2000) \citep{hamraoui2000can} have studied the kinetics of capillary rise of pure water and pure ethanol as well as their mixtures that, under static conditions, wet glass capillary tubes in both dry and prewetting wall conditions. Specifically, they have postulated a dynamic contact angle term that is linearly dependent on the velocity of the capillary rise and whose correction, in this linear approximation, takes on the form of a three-phase line friction coefficient, $M$, equivalent to our parameter $\alpha$, up to a proper dimensionalization factor. The value of $M$ for ethanol, water and a water-ethanol mixture is reported in table~\ref{tab:C10_FluidProp}.
\begin{table}
\centering
\begin{tabular}{c|c|c|c|c|c|c}
\hline
liquid & $\rho\left(\text{kg/m$^3$}\right)$ & $\gamma\left(\text{N/m}\right)$ & $\nu\left(\text{m$^2$/s}\right)$ & $M\left(\text{Pa s}\right)$ & $\bar{\alpha}=\frac{M}{\gamma}\left(\text{s/m}\right)$ & $\alpha=\bar{\alpha}\frac{\gamma}{\nu\rho}\,\left(\text{rad}\right)$ \\ \hline
water & 1000 & 0.072 & 1.0$\times$10$^{-6}$ & 0.2 & 6.25 & 200\\
mixture & 983 & 0.050 & 1.0$\times$10$^{-6}$ & 0.14 & 2.8 & 140\\
ethanol & 786 & 0.022 & 1.4$\times$10$^{-6}$ & 0.04 & 1.82 & 36\\ \hline
\end{tabular}
\caption{Value of the non-dimensional contact line parameter $\alpha$ for water, water-ethanol mixture and pure ethanol as measured by Hamraoui \textit{et al.} (2000) \citep{hamraoui2000can}. The dimensional value of the friction coefficient $M$ (denoted by $\beta$ in their study) is here converted in the dimensional, $\overline{\alpha}$, and non-dimensional, $\alpha$, contact line parameter.}
\label{tab:C10_FluidProp}
\end{table}\\
\indent Particularly relevant to our study is the value measured by Hamraoui \textit{et al.} (2000) for pure water, $M=0.2\,\text{Pa s}$, which translates into $\alpha=200\,\text{rad}$, hence matching precisely the value found to fit the experimental data. As a side comment, the use of the coefficient $\alpha$ also produces an increase in the natural frequencies, thus bringing the numerics closer to the experimental values.\\
\indent Through this careful comparison with experiments by Hamraoui \textit{et al.} (2000) and Dollet \textit{et al.} (2020), we have been capable of quantifying numerically the natural properties of the system in the two dynamical phases of interest, handled independently. All our estimates and hypotheses seem consistent with these measurements.\\
\indent The idea is now to combine the two separated descriptions for the pinned-phase and free-phase, so as to account for a dynamic change in the contact line boundary conditions and predict the nonlinear relaxation dynamics. This is done in the next section by employing the projection algorithm.


\bigskip
\begin{centering}\section{Projection method}\label{sec:C10_Sec3}\end{centering}

\begin{figure}[]
\centering
\includegraphics[width=1\textwidth]{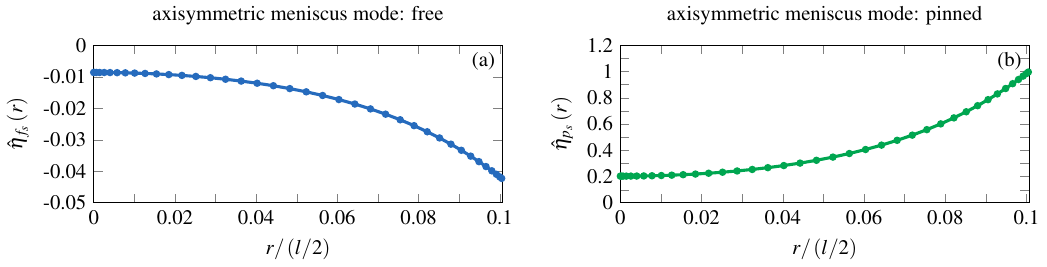}\\
\bigskip
\includegraphics[width=1\textwidth]{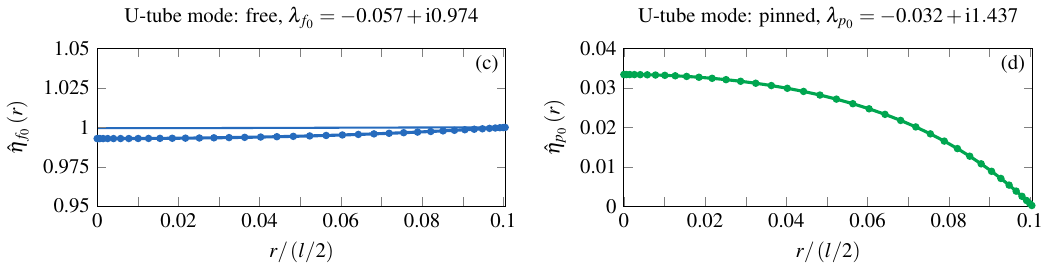}\\
\bigskip
\includegraphics[width=1\textwidth]{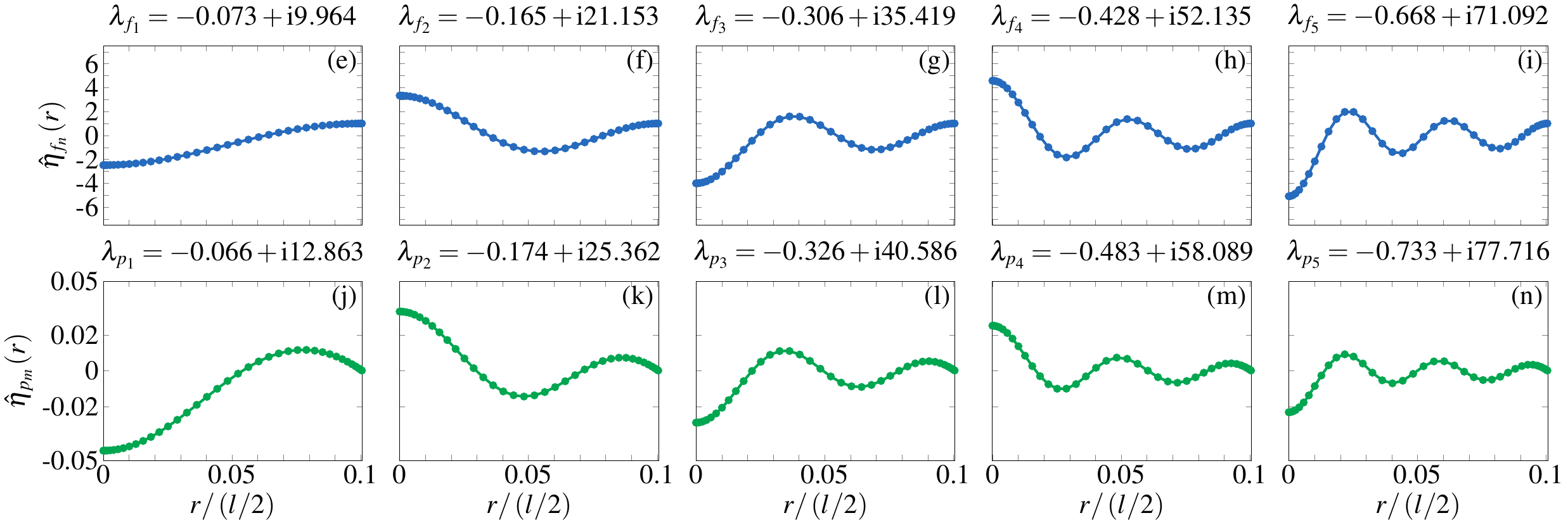}
\caption{(a) Axisymmetric meniscus modes associated with the free-phase and (b) with the pinned-phase. In (a), \textcolor{black}{the slope at the wall is 1}, whereas the contact line elevation is $F_0$. In (b), the \textcolor{black}{slope} is $1/F_0$, whereas the contact line elevation is 1. (c) Real part of the eigen-interface associated with the free and (d) pinned U-tube modes, with the corresponding eigenvalues, $\lambda_{f_0}=-\sigma_{f_0}+\text{i}\omega_{f_0}$ and $\lambda_{p_0}=-\sigma_{p_0}+\text{i}\omega_{p_0}$ reported on top. The free mode is normalized such that the contact line elevation is 1, while the pinned mode is normalized such that the slope at the wall is 1. For completeness, in (c), we have also reported the interface shape when $\alpha=0$ (thin blue line) as shown in Fig.~\ref{fig:C10_FIg17bis}(a). (e)-(i) Real part of the eigen-interface associated with the five least damped free and (j)-(n) pinned capillary-gravity waves. The same normalization as in (c) and (d) is employed.}
\label{fig:C10_AllModes}
\end{figure}

\begin{centering}\subsection{General formalism}\end{centering}

A detailed step-by-step description of the projection algorithm is already provided in Bongarzone \textit{et al.} (2021) \citep{bongarzone2021relaxation}. In this section, we recall the salient points of the method and we comment on the few differences intrinsic to specific dynamics of the problem here considered.\\
\indent When the contact line motion is schematized using Hocking’s law amended with a static hysteresis range, we can identify two well-distinct phases of the dynamics, one in which the angle varies linearly with a slope $\alpha$ as a function of the contact line speed, $Ca\partial\eta/\partial t$ (Hocking’s linear law) and one in which the contact line is pinned at a certain elevation with zero velocity (static hysteresis) and the angle changes from $\theta_s+\theta^+$ to $\theta_s+\theta^-$ ($\Delta=\theta^+-\theta^-)$ or vice versa. We remind that we denote these two phases as free, $_{f}$, and pinned, $_{p}$, phase, respectively.\\
\indent The solution in these two phases is then expressed as the sum of the corresponding particular static solution (meniscus mode), $\mathbf{q}_{f_s}$ and $\mathbf{q}_{p_s}$ (the subscripts $_{f_s,p_s}$ stand for free-static or pinned-static), and a truncated basis of linear eigenmodes, $\hat{\mathbf{q}}_{f_n}$ and $\hat{\mathbf{q}}_{p_m}$, weighted by their unknown amplitudes:
\begin{subequations}
\begin{equation}
\label{eq:C10_projF}
\mathbf{q}_f=\underbrace{\theta^{\pm}\mathbf{q}_{f_s}}_{\text{free-end meniscus mode}}+\underbrace{\left(A_0\hat{\mathbf{q}}_{f_0}e^{\lambda_{f_0}\left(t-T_f\right)}+c.c.\right)}_{\text{free-end U-tube mode}}+\underbrace{\left(\sum_{n=1}^{N_f}A_{f_n}\hat{\mathbf{q}}_{f_n}e^{\lambda_{f_n}\left(t-T_f\right)}+c.c.\right)}_{\text{free-end capillary-gravity waves}}
\end{equation}
\begin{equation}
\label{eq:C10_projP}
\mathbf{q}_p=\underbrace{e_{fp}\mathbf{q}_{p_s}}_{\text{pinned-end meniscus mode}}+\underbrace{\left(B_0\hat{\mathbf{q}}_{p_0}e^{\lambda_{p_0}\left(t-T_p\right)}+c.c.\right)}_{\text{pinned-end  U-tube mode}}+\underbrace{\left(\sum_{m=1}^{M_p}B_{p_m}\hat{\mathbf{q}}_{p_m}e^{\lambda_{p_m}\left(t-T_p\right)}+c.c.\right)}_{\text{pinned-end capillary-gravity waves}}
\end{equation}
\end{subequations}
\noindent All these ingredients are visually summarized in Fig.~\ref{fig:C10_AllModes}. As described in the previous section and in contradistinction with the two-dimensional system of Ref.~\onlinecite{bongarzone2021relaxation}, the present U-tube dynamics is characterized by two families of oscillating natural modes, namely a free/pinned U-tube mode ($n=0$ or $m=0$) and free/pinned capillary-gravity waves ($n \in \left[1,N_f\right]$, $m \in \left[1,M_p\right]$). However, these waves oscillate at a much larger frequency and are more damped than the U-tube modes. Accounting for them in the algorithm is useful if one is interested in capturing fast transients, but with the purpose of modelling the global dynamical features of the system, their inclusion in the analysis is not strictly necessary. Hereinafter we will ignore the capillary-gravity waves, and we will only retain the dominant free and pinned U-tube natural modes described in \S\ref{sec:C10_Sec2} and here denoted by $\hat{\mathbf{q}}_{f_0}$ (free) and $\hat{\mathbf{q}}_{p_0}$ (pinned), with amplitudes $A_0$ and $B_0$, and eigenvalues $\lambda_{f_0}=-\sigma_{f_0}+\text{i}\omega_{f_0}$ and $\lambda_{p_0}=-\sigma_{p_0}+\text{i}\omega_{p_0}$, respectively.\\
\indent Including a meniscus mode in the solution form~\eqref{eq:C10_projF} associated with the free-phase, i.e. $\mathbf{q}_{f_s}$, is necessary in order to properly deal with the non-homogeneous term in the right-hand-side of the contact line condition~\eqref{eq:C10_free_bc}. The particular solution resulting from this static forcing term, $-\theta^{\pm}$, consists in a static meniscus modification $\eta_{f_s}$ (with $\mathbf{u}_{f_s}=\mathbf{0}$) that satisfies the linearized meniscus equation
\begin{equation}
\label{eq:C10_MenModF}
\eta_{f_s}-\frac{1}{Bo}\left[\frac{1}{\left(1+\eta_{0,r}^2\right)^{3/2}}\frac{\partial^2\eta_{f_s}}{\partial r^2}+\frac{\left(1+\eta_{0,r}^2\right)}{\left(1+\eta_{0,r}^2\right)^{5/2}}\frac{1}{r}\frac{\partial \eta_{f_s}}{\partial r}\right]=0,\ \text{with}\ \left.\frac{\partial \eta_{f_s}}{\partial r}\right|_{r=a/\left(l/2\right)}=-\theta^{\pm},
\end{equation}
\noindent with the terms in brackets representing the first-order variation of the nonlinear curvature linearized around the static meniscus $\eta_0$ and applied to $\eta_{f_s}$. For the convenience of notation, note that, in Eq.~\eqref{eq:C10_MenModF}, we actually impose \textcolor{black}{the slope} $\partial\eta_{f_s}/\partial r=-1$ instead of $-\theta^{\pm}$, while keeping the term $\theta^{\pm}$ explicit in front of the particular solution in~\eqref{eq:C10_projF}.\\
\indent The pinned-condition~\eqref{eq:C10_pinn_bc} is homogeneous and it is explicitly accounted for in the corresponding eigenvalue problem. However, the condition $\partial\eta/\partial t=0$ also allows for a static particular solution with $\eta_{p_s}=\text{constant}$ at the contact line $r=a/\left(l/2\right)$ (and with $\mathbf{u}_{p_s}=\mathbf{0}$). The meniscus mode for the pinned-phase is therefore computed as $\eta_{p_s}=\eta_{f_s}/F_0$, with $F_0$ the value of $\eta_{f_s}$ at the wall $r=a/\left(l/2\right)$, so as to have a unitary value, $\eta_{p_s}=1$, at $r=a/\left(l/2\right)$ (see Fig.~\ref{fig:C10_AllModes}). This unitary value is weighted by the contact line elevation $e_{fp}$ in~\eqref{eq:C10_projP}, with $e_{fp}$ kept fixed during the pinned-phase.

\bigskip
\begin{centering}\subsection{Workflow of the method}\end{centering}

\begin{figure}[]
\centering
\includegraphics[width=1\textwidth]{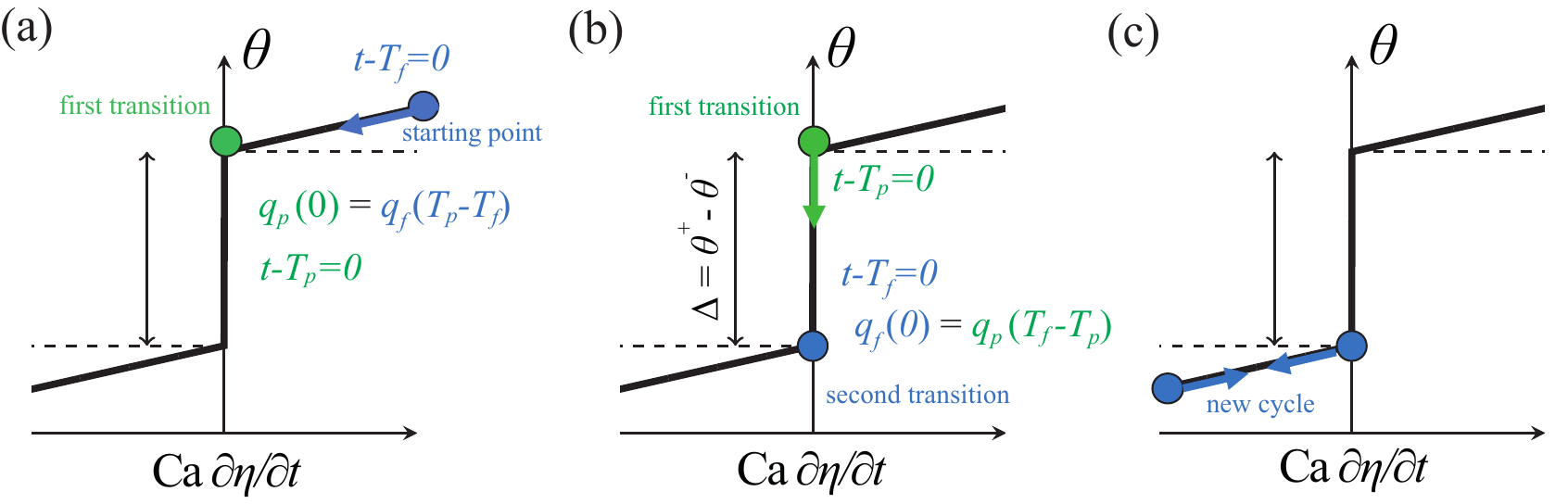}
\caption{Workflow of the projection algorithm (from (a) to (c)).}
\label{fig:C10_StepProj}
\end{figure}

A visual workflow of the algorithm is illustrated in Fig.~\ref{fig:C10_StepProj}. Let us suppose to initialize the system in the upper free-phase (panel (a)) by assigning the amplitude of the free U-tube mode, $A_0$, at $t-T_f=0$. The system is let evolve in time according to~\eqref{eq:C10_projF}. When the contact line speed reaches the null value, we have the first transition, i.e. from free to pinned. At this time instant, $t=T_p$, we require the continuity of all variables of the system, i.e. $\mathbf{q}_{p}\left(0\right)=\mathbf{q}_{f}\left(T_p-T_f\right)$. This corresponds to imposing
\begin{equation}
\label{eq:C10_PROJ1}
\theta^+\mathbf{q}_{f_s}+\left(A_0\hat{\mathbf{q}}_{f_0}e^{\left(-\sigma_{f_0}+\text{i}\omega_{f_0}\right)\left(T_p-T_f\right)}+c.c.\right)=e_{fp}\mathbf{q}_{p_s}+\left(B_0\hat{\mathbf{q}}_{p_0}+c.c.\right),
\end{equation}
\noindent which, using the fact that the contact line elevation at the end of the free-phase reads (noting that $\hat{\eta}_{f_0}=1$ at $r=a/\left(l/2\right)$ and $\eta_{p_s}=\eta_{f_s}/F_0$)
\begin{equation}
\label{eq:C10_PROJ2}
e_{fp}=\theta^+F_0+\left(A_0\,e^{\left(-\sigma_{f_0}+\text{i}\omega_{f_0}\right)\left(T_p-T_f\right)}+c.c.\right),
\end{equation}
\noindent can be conveniently rewritten as
\begin{equation}
\label{eq:C10_PROJ3}
B_0\hat{\mathbf{q}}_{p_0}+c.c.=A_0\left(\hat{\mathbf{q}}_{f_0}-\mathbf{q}_{p_s}\right)e^{\left(-\sigma_{f_0}+\text{i}\omega_{f_0}\right)\left(T_p-T_f\right)}+c.c.\equiv\mathbf{f}_{fp},
\end{equation}
\noindent where the resulting term on the right-hand side is fully known.\\
\indent The amplitude of the U-tube mode pertaining to the next pinned-phase, $B_0$, still unknown at this stage, is computed by projecting, with respect to a specific weighted inner product, the final-time free solution, $\mathbf{f}_{fp}$, on the initial-time pinned solution as
\begin{equation}
\label{eq:C10_PROJ4}
B_0=<\hat{\mathbf{q}}_{p_0}^{\dagger},\mathbf{f}_{fp}>_E.
\end{equation}
\noindent with $\hat{\mathbf{q}}_{p_0}^{\dagger}$ the adjoint U-tube pinned-mode.\\
\indent We are now entering the pinned-phase (panel (b)). The initial contact angle is $\theta_s+\Delta/2=\theta_s+\theta^+$, and the time-evolution of the system is described by~\eqref{eq:C10_projP}. The contact angle progressively changes with a fixed contact line elevation $e_{fp}$ and once it reaches the value $\theta_s-\Delta/2=\theta_s+\theta^-$, the second transition occurs. We impose again the continuity of the flow variables, i.e. $\mathbf{q}_{f}\left(0\right)=\mathbf{q}_{p}\left(T_f-T_p\right)$, 
\begin{equation}
\label{eq:C10_PROJ5}
e_{fp}\mathbf{q}_{p_s}+\left(B_0\hat{\mathbf{q}}_{p_0}e^{\left(-\sigma_{p_0}+\text{i}\omega_{p_0}\right)\left(T_f-T_p\right)}+c.c.\right)=\theta^-\mathbf{q}_{f_s}+\left(A_0\hat{\mathbf{q}}_{f_0}+c.c.\right),
\end{equation}
\noindent with 
\begin{equation}
\label{eq:C10_PROJ6}
\theta^-=e_{fp}/F_0+\left(B_0\,e^{\left(-\sigma_{p_0}+\text{i}\omega_{p_0}\right)\left(T_f-T_p\right)}+c.c.\right),
\end{equation}
\noindent so that Eq.~\eqref{eq:C10_PROJ5} can be rearranged as
\begin{equation}
\label{eq:C10_PROJ7}
A_0\hat{\mathbf{q}}_{f_0}+c.c.=B_0\left(\hat{\mathbf{q}}_{p_0}-\mathbf{q}_{f_s}\right)e^{\left(-\sigma_{p_0}+\text{i}\omega_{p_0}\right)\left(T_f-T_p\right)}+c.c.\equiv\mathbf{f}_{pf}.
\end{equation}
\noindent We thus project the final-time pinned solution on the initial-time free solution, so as to determine the new amplitude $A_0$. 
\begin{equation}
\label{eq:C10_PROJ8}
A_0=<\hat{\mathbf{q}}_{f_0}^{\dagger},\mathbf{f}_{pf}>_E.
\end{equation}
\noindent with $\hat{\mathbf{q}}_{f_0}^{\dagger}$ the adjoint U-tube free-mode.\\
\indent The system enters the lower free-phase (panel (c)) and the cycle is repeated over again. Each projection eventually induces a rapid loss of total energy in the liquid motion and contributes to its nonlinear damping. After a few cycles, the inertia of the oscillating liquid column will no longer be sufficient to surpass the static solid-like friction and the system will get trapped in the pinned-phase. The secondary fluid bulk motion following the arrest of the contact line will decay exponentially under the effect of the linear viscous dissipation characteristic of the pinned dynamics.

\bigskip
\begin{centering}\subsection{$E$-norm inner product and definition of adjoint modes}\end{centering}

We note that, owing to the axisymmetric configuration, the inner product employed in this context differs from that used in Ref.~\onlinecite{bongarzone2021relaxation}:
\begin{equation}
\label{eq:C10_scalar_prod}
<\mathbf{w},\mathbf{u}>_E=\int_{V}\overline{\mathbf{u}}_{\mathbf{w}}\mathbf{u}_{\mathbf{v}}\,r\text{d}r\text{d}z+\int_{z=\eta_0\left(r\right)}\left[\overline{\eta}_{\mathbf{w}}\eta_{\mathbf{v}}+\frac{1}{Bo}\left(\frac{1}{\left(1+\eta_{0,r}^2\right)^{3/2}}\right)\frac{\partial\overline{\eta}_{\mathbf{w}}}{\partial r}\frac{\partial\eta_{\mathbf{v}}}{\partial r}\right]\,r\text{d}r
\end{equation}
\noindent where $\mathbf{v}=\left\{\mathbf{u}_{\mathbf{v}},p_{\mathbf{v}},\eta_{\mathbf{v}}\right\}^T$ and $\mathbf{w}=\left\{\mathbf{u}_{\mathbf{w}},p_{\mathbf{w}},\eta_{\mathbf{w}}\right\}^T$ are two generic vectors, the bar designates the complex conjugate and the subscript $_{E}$ stands for \textit{energy}. We recall that~\eqref{eq:C10_scalar_prod} represents the total energy norm, where the volume integral measures the kinetic energy, whereas the two boundary terms are, respectively, the gravitational and surface potential energies. We also note that the surface integral associated with the surface energy (curvature term) is further weighted by $\left(1+\eta_{0,r}^2\right)^{-3/2}$, resulting from the linearization around an initially curved static meniscus, $\eta_0\left(r\right)\ne0$.\\
\indent As a final comment, in Eqs.~\eqref{eq:C10_PROJ4}-\eqref{eq:C10_PROJ8} we have invoked the concept of adjoint modes, solutions of the adjoint linearized homogeneous problem, whose formal derivation is given in the supplementary material of Bongarzone \textit{et al.} (2021) \citep{bongarzone2021relaxation}. In this regard, here we limit ourselves to reporting the final result, according to which
\begin{equation}
\label{eq:C10_ADJ}
\hat{\mathbf{q}}_{f,p}^{\dagger}=\left\{
\begin{matrix}
\hat{\mathbf{u}}^{\dagger}\\
\hat{p}^{\dagger}\\
\hat{\eta}^{\dagger}
\end{matrix}\right\}_{f,p}=\left\{
\begin{matrix}
-\overline{\hat{\mathbf{u}}}\\
-\overline{\hat{p}}\\
\overline{\hat{\eta}}
\end{matrix}\right\}_{f,p}\ne\hat{\mathbf{q}}_{f,p}, \ \ \ \ \ \ \ \ \lambda_{f,p}^{\dagger}=-\sigma_{f,p}-\text{i}\omega_{f,p}=\overline{\lambda}_{f,p}.
\end{equation}
\noindent The abovementioned supplementary notes also provide a demonstration that direct modes, $\hat{\mathbf{q}}_{f,p}$ and adjoint modes, $\hat{\mathbf{q}}_{f,p}$, form a bi-orthogonal basis with respect to the scalar product~\eqref{eq:C10_scalar_prod}, with the adjoint modes that appear, therefore, as the most suitable choice for the projection step.

\bigskip
\begin{centering}\section{Comparison with experiments and results}\label{sec:C10_Sec4}\end{centering}

\begin{centering}\subsection{Contact line dynamics and finite-time arrest}\end{centering}

\begin{figure}[]
\centering
\includegraphics[width=1\textwidth]{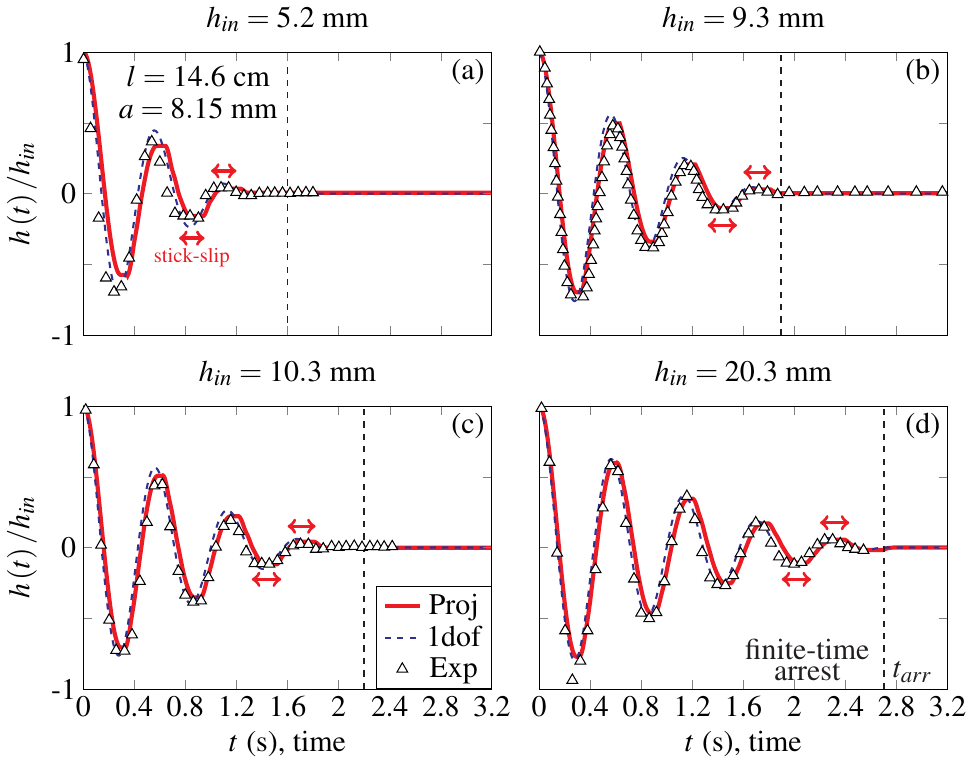}
\caption{Contact line elevation versus time for different initial conditions. Dashed line: 1dof model. Red solid lines: predictions from the projection model. Markers: experiments by Dollet \textit{et al.} (2020). We note that in performing the calculation, we have actually considered an effective tube length of $16.2\,\text{cm}$, where an excess length of $l'=1.6\,\text{cm}$ is introduced in order to into account the fact that the cross-section along the curved part of the tube is not constant due to the fabrication process. See Ref.~\onlinecite{dollet2020transition} for further details.}
\label{fig:C10_Fig5}
\end{figure}

\begin{figure}[]
\centering
\includegraphics[width=0.65\textwidth]{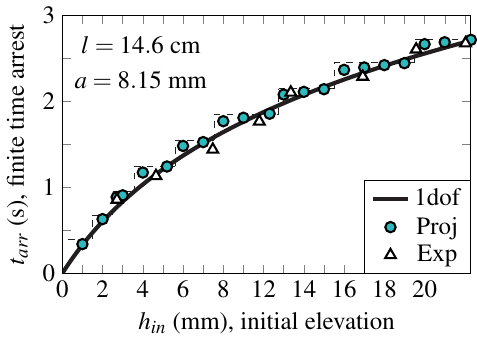}
\caption{Finite time of arrest versus the imposed initial elevation. Black solid line: analytical prediction from the one-degree-of-freedom model proposed by Dollet \textit{et al.} (2020). White triangles: experimental measurements by Dollet \textit{et al.} (2020). Colored circles: projection method. The black dashed line only serves to guide the eyes.}
\label{fig:C10_Fig6}
\end{figure}

In this section, the most relevant results are discussed. First, we compare the contact line dynamics predicted by the projection method versus that predicted by the 1dof model and that measured experimentally by Dollet \textit{et al.} (2020) \citep{dollet2020transition}. This comparison is outlined in Fig.~\ref{fig:C10_Fig5} for different initial contact line elevations, $h_{in}$. The improvement brought by the present projection method is not striking from this comparison. Both the 1dof model and the present model are in fairly good agreement with experiments. Nevertheless, we can spot, e.g. in panels (a,b,c), that our model seems to capture the stick-slip transitions preceding the contact line arrest. Those transitions are visible in the experiments and correspond to the dynamical phases where the contact line elevation remains approximately constant over a time interval, as indicated by the red arrows.
\\
\indent An interesting aspect highlighted by the projection model is related to the dependence of the finite-time arrest for the contact line, $t_{arr}$, on the initial elevation, $h_{in}$. The time arrest of the contact line is indicated in Fig.~\ref{fig:C10_Fig5} by the vertical black dashed lines, while its dependence on $h_{in}$ is characterized more in detail in Fig.~\ref{fig:C10_Fig6}, which shows how $t_{arr}$ follows a step-like function.  
\\
\indent From our knowledge, such a trend has not been reported in the literature yet, but it appears intuitively correct. Indeed, the arrest of the contact line occurs when, after a few oscillation cycles, the inertia of the system is no longer sufficient to overcome this static friction. Fig.~\ref{fig:C10_Fig6} suggests that there are ranges of initial elevations $h_{in}$ for which the final time of arrest is $t_{arr}$ remains unchanged. In order to prolongate in time the oscillatory contact line motion, the system needs to surpass this final energy barrier, which is only possible by starting from a sufficiently larger potential energy, and thus, from a larger $h_{in}$.

\bigskip
\begin{centering}\subsection{Global damping properties and frequency modulation}\end{centering}

\begin{figure}[]
\centering
\includegraphics[width=0.8\textwidth]{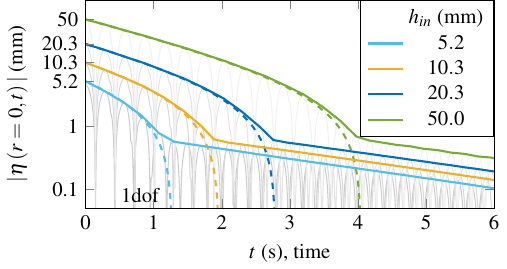}
\caption{Centerline free surface elevation, i.e. $r=0$ and $z=0$ (in log-scale), versus time for different initial elevations, $h_{in}$. The grey solid lines show the actual signal produced by the projection method, while the coloured solid lines indicate the amplitude envelope only. The coloured dashed lines correspond to the analytical prediction given by the single-degree-of-freedom model employed by Dollet \textit{et al.} (2020). An almost abrupt change in the trend of these signals is well visible. This is a clear sign of the final transition to a pinned contact line dynamics following the contact line arrest.}
\label{fig:C10_Fig16}
\end{figure}

As the projection method deals with the full hydrodynamic system, we have access to all the degrees of freedom of the system. Looking away from the contact line and rather focusing the attention, for example, on the centerline dynamics at $r=0$, the useful insights brought by the present approach are evident. The centerline dynamics is of course affected by what happens at the contact line, but at the same time, it does not undergo a finite-time arrest. The associated time series, computed for different initial elevations, is reported in Fig.~\ref{fig:C10_Fig16}.
\\
\indent An inspection of this time-signal evolution reveals, consistently with previous experimental observations \citep{Cocciaro93}, how the contact line arrest is followed by the secondary bulk motion characterized by an exponential relaxation with a constant damping coefficient (i.e. the final linear trend in the log-scale plot of Fig.~\ref{fig:C10_Fig16}), which is completely overlooked by the 1dof model. By monitoring the nonlinear decay of such a signal, we can estimate the damping rate and the modulation of the oscillation frequency as a function of the time-dependent oscillation amplitude
\begin{figure}[]
\centering
\includegraphics[width=0.8\textwidth]{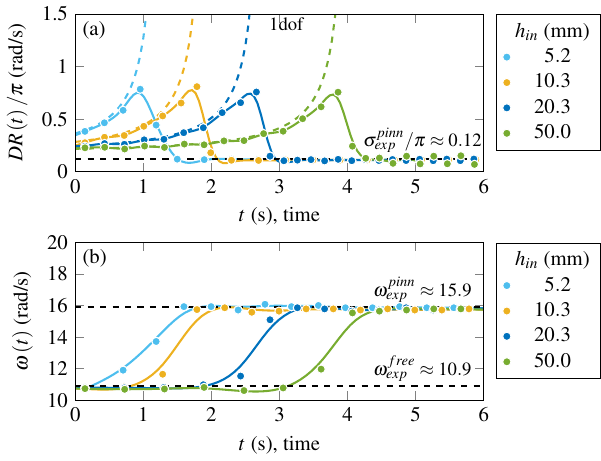}
\caption{(a) Dimensional damping rate and (b) frequency modulation versus time at different initial conditions. The damping rate, $DR\left(t\right)$ is computed as the logarithmic decrement of the amplitude of the centerline free surface elevation, shown in Fig.~\ref{fig:C10_Fig16}. The frequency is computed from the same signal by evaluating the period from peak to peak, with the resulting value that is then roughly assigned to the midpoint of the corresponding time interval (coloured filled circles in (a) and (b)). The coloured solid lines represent the best fit \textcolor{black}{(smoothing splines)} of these time signals, whereas the coloured dashed lines correspond to the analytical prediction given by the single-degree-of-freedom model employed by Dollet \textit{et al.} (2020).}
\label{fig:C10_Fig15}
\end{figure}\\
\indent The result of this procedure is explained and illustrated in Fig.~\ref{fig:C10_Fig15}. Similarly to the weakly nonlinear analysis formalized by Viola \& Gallaire (2018) \citep{Viola2018b}, the 1dof model predicts the initial increase in the damping rate, $DR\left(t\right)$, but it diverges around $t\approx t_{arr}$. This finite-time singularity is not surprising as the contact line arrests at $t=t_{arr}$, but it is only locally correct, and it does not represent a good description of the global damping rate. On the contrary, the damping rate resulting from the projection shows an increase as the wave amplitude decreases, until it reaches a maximum value, at a time instant close to $t=t_{arr}$ after which it decreases to a nearly constant value. Once the pinned dynamics is established, the damping rate is approximately constant and equal to the viscous damping coefficient of the pinned U-tube mode. Concerning the frequency modulation in time, we find a smooth evolution from the characteristic value of the initially dominant free U-tube mode to a final value, reached for $t\approx t_{arr}$ and corresponding to the natural oscillation frequency of the pinned U-tube mode. Although no results concerning the damping rate and frequency modulation in time were reported in Ref.~\onlinecite{dollet2020transition}, the initial and final values match well the experimental ones (as indicated in Fig.~\ref{fig:C10_Fig15} by the values of $\omega_{exp}^{free}$, $\omega_{exp}^{pinn}$ and $\sigma_{exp}^{pinn}$), and the intermediate behaviour is fully consistent with that experimentally reported by Cocciaro \textit{et al.} (1993) \citep{Cocciaro93} in a sloshing configuration.\\
\indent We note that the centerline elevation, as the contact line elevation, is also a local measurement, but it is more representative of the overall dynamics. Similar trends for the damping and frequency are found by monitoring, e.g., the decay of the total energy (see Ref.~\onlinecite{bongarzone2021relaxation}), which represents instead a global observable.

\bigskip
\begin{centering}\section{Conclusions}\label{sec:C10_Conclusion}\end{centering}

In this work, we have employed the projection method developed in Bongarzone \textit{et al.} (2021) \citep{bongarzone2021relaxation} to study the natural relaxation dynamic of small amplitude liquid oscillations in a U-shaped tube, as experimentally investigated by Dollet \textit{et al.} (2020) \citep{dollet2020transition}.\\
\indent First, we attempted to rationalize the linear dissipation properties of the system in both the free and pinned dynamical phases so as to explain the fitting parameter used in the 1dof model of Dollet \textit{et al.} (2020) (see Eq.~\eqref{eq:C10_1dof}). After having numerically estimated the effect of three-dimensionality, i.e. of the tube curvature, and the contribution of the Stokes boundary layers on the overall linear damping coefficients (see Appendices~\ref{sec:C10_AppA} and~\ref{sec:C10_AppB}), a linear Hocking's law for the dynamic variation of the contact angle with the contact line speed has been accounted for in order to compensate for the missing dissipation, hence allowing for a good match with experiments. The combination of such a linear law with the static hysteresis range considered in Dollet \textit{et al.} (2020) translates into the phenomenological nonlinear contact line model already used in Refs.~\onlinecite{Hocking87,hamraoui2000can,bongarzone2021relaxation}.\\
\indent The full hydrodynamic system, supplemented with this contact line model, has been then studied in the framework of the projection approach, so as to compare the resulting predictions with those from the simple 1dof damped pendulum model employed in Dollet \textit{et al.} (2020) and with their experimental measurements. When looking at the contact line dynamics only, the improvement brought by the present model is not striking. Both the 1dof model and the present model are in fairly good agreement with experiments and predict well the contact line arrest. However, our model seems to correctly capture some of the stick-slip transitions occurring, in a more pronounced way, just before the finite-time arrest. If one is interested in having a quick estimation of the finite-time arrest for the contact line, we, therefore, recommend using the damped pendulum model.\\
\indent Nevertheless, although the peculiar contact line dynamics, with its stick-slip motion and finite-time arrest, is the main responsible for the initial nonlinear dissipation of the system, it is not fully representative of the global dynamics. Through the projection method, we have access to all the degrees of freedom of the system. This allowed us to explore, for example, the centerline dynamics, which is affected by what happens at the contact line but does not undergo a finite-time arrest. An inspection of this time-signal evolution reveals, consistently with previous experimental observations \citep{Cocciaro93} in the context of sloshing dynamics, how the contact line arrest is followed by the secondary bulk motion characterized by an exponential relaxation. By monitoring the nonlinear decay of such a signal obtained via the projection approach, we have been able to estimate the damping rate and the oscillation frequency (both amplitude-dependent) of the system, hence correctly capturing the transition from an initial stick-slip motion to a final pinned dynamics, which has been so far overlooked by the theoretical analyses reported in the literature.\\
\indent The projection method, here applied to the case of a piecewise linear contact line model, has already been generalized to any smooth non-linear contact line dynamics, e.g. a cubic law according to the \textit{Dussan} model (see Ref.~\onlinecite{bongarzone2021relaxation}). Replacing the linear Hocking's law with a more sophisticated nonlinear law, e.g. cubic, and combining the latter with a range of static hysteresis is of interest and appears natural. Other future perspectives include the introduction into the model of small amplitude external forcing, i.e. axial time-harmonic excitations, and the extension to three-dimensional non-axisymmetric oscillatory dynamics, which is of great relevance for sloshing-related problems \citep{bongarzone2022amplitude,marcotte2023super,marcotte2023swirling} and in the description of oscillatory sessile drop dynamics \citep{noblin2004vibrated,xia2018moving,amberg2022detailed,ludwicki2022contact}.\\

\bigskip
\begin{centering}\section*{Acknowledgments}\end{centering}
\indent We acknowledge the financial support of the Swiss National Science Foundation under grant 178971. We also acknowledge Bastien Ravot for fruitful discussions on slip length models.\\

\indent The authors declare the absence of any conflict of interest.

\bigskip
\appendix

\begin{centering}\section{Effect of the tube curvature on the damping}\label{sec:C10_AppA}\end{centering}

\begin{figure}[]
\centering
\includegraphics[width=0.8\textwidth]{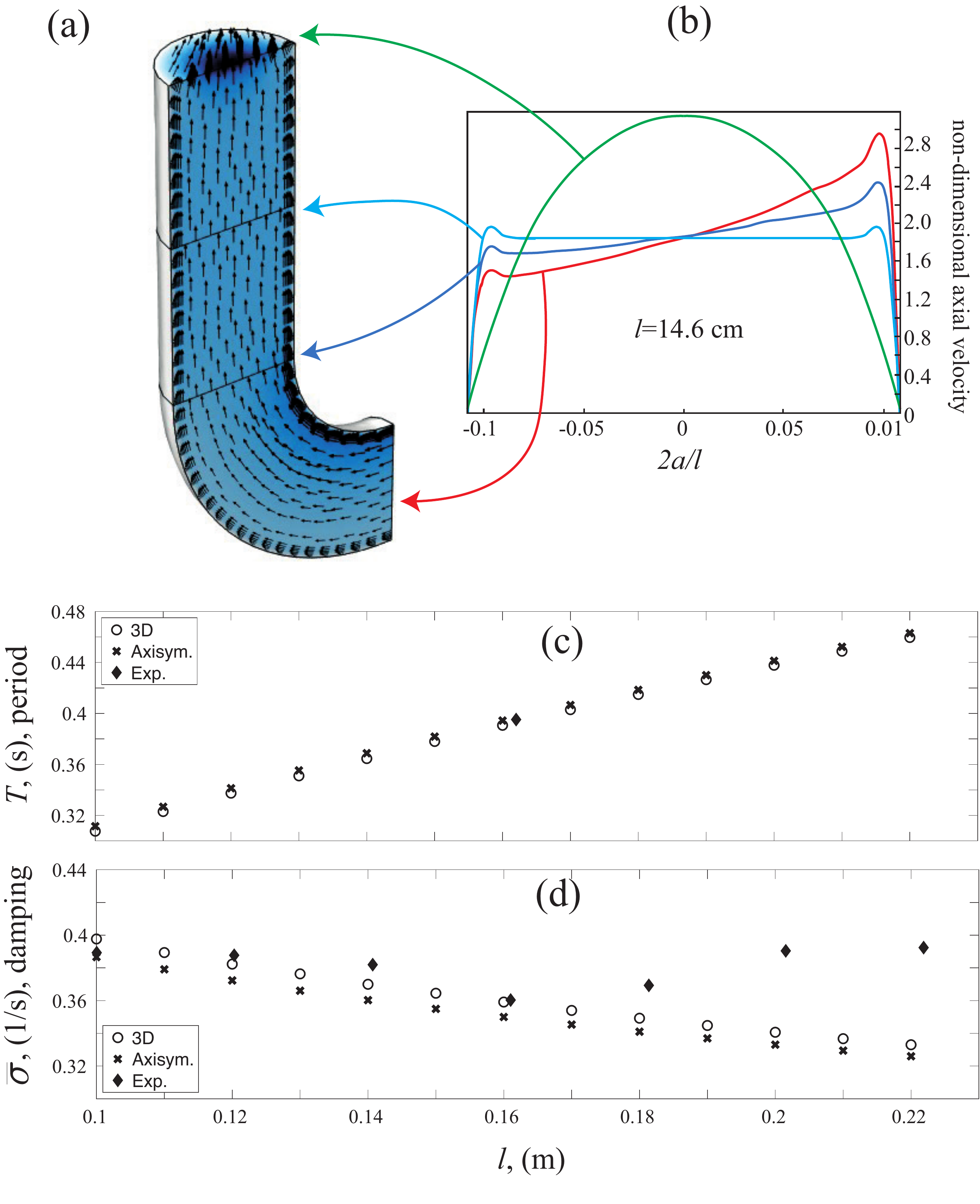}
\caption{(a) Three-dimensional natural U-tube mode for a pinned contact line. The full domain has been resolved, but only a quarter of it is shown here for visualization purposes. (b) Axial velocity profile plotted at different sections along the tube, as indicated by the coloured arrows. The liquid column length in (a) and (b) has been set to $l=14.6\,\text{cm}$. (c) Dimensional oscillation period, $T=2\pi/\omega$, associated with the pinned contact line dynamics and as a function of the liquid column length, $l$. (d) Same as in (c), but for the dimensional damping coefficient. In (c) and (d), empty circles correspond to the present 3D calculation, black crosses are from the axisymmetric model discussed throughout the manuscript, while filled black diamonds are experimental measurements from Ref.~\onlinecite{dollet2020transition}. Only one measurement has been reported for the oscillation period.}
\label{fig:C10_3Dstab}
\end{figure}

In this Appendix, we perform the full three-dimensional eigenvalue analysis for a pinned contact line. The latter condition is easier to resolve numerically, as no stress singularity emerges from the imposition of a no-slip wall. Although the flow dynamics for a moving contact line and the resulting damping properties may differ from the one considered here, the purpose of this appendix is simply to have a first estimation of the effect of the curved part of the tube on the global linear damping coefficient. This computation serves us to partially justify the fundamental assumption of neglecting the tube curvature. With respect to the real experiment, we can only obtain a rough estimation, as the tube used by Dollet \textit{et al.} (2020) \citep{dollet2020transition} shows a significantly smaller cross-section in its curved part than in its straight parts, where it is circular of uniform radius $a=8.15\,\text{mm}$ within a few tens of microns. As it is difficult to measure this variation locally, \textcolor{black}{we lack information to mesh numerically the actual geometry with full fidelity}. For these reasons, we will simply consider a constant cross-section of radius $a$.\\
\indent Thus, the linearized governing equations with their boundary conditions have been implemented in the finite-element software COMSOL Multiphysics v5.6. To mesh the physical domain, we have adopted a hybrid \textcolor{black}{hexahedrical-tetrahedrical} mesh. Specifically, \textcolor{black}{tetrahedral} elements were used in the interior, while \textcolor{black}{hexahedron} elements were adopted in the neighbourhood of the free surface, sidewalls and bottom, where, in addition, boundary layer refinements were used to better model the viscous Stokes boundary layers. The linearized equations were manually written in their weak formulation using the Weak Form PDE tools available in the software. We used P2 for the velocity field and P1 elements for the pressure field, so as to avoid spurious pressure mode. The interface variable was discretized with P2 elements. Globally, the grid is made of approximately 300\,000 degrees of freedom, for which convergence was tested.\\
\indent The results of this computation are reported in Fig.~\ref{fig:C10_3Dstab}. Panel (a), gives a picture of the three-dimensional natural U-tube mode for a pinned contact line: the full domain has been resolved, but for visualization purposes, only a quarter of it is shown. The non-dimensional axial velocity profile is reported in panel (b) at different locations along the tube as indicated by the colored arrows. We can see how the effect of the curvature is locally important from the asymmetry in the velocity profile: the velocity is higher where the curvature is higher. This asymmetric profile gradually adapts to a symmetric plug-like flow in the straight arm of the tube, and eventually, it relaxes to a bell-like profile at the interface. This last profile seems peculiar, but it is consistent with the fact that the axial velocity at the surface equals the time derivative of the interface, which, for a pinned dynamics, has indeed a bell-like shape (see \S\ref{sec:C10_Sec2}).\\
\indent Although the curvature seems to affect the flow locally, Fig.~\ref{fig:C10_3Dstab}(c) and (d) suggest that it does not significantly influence the eigenvalue properties of the system, i.e. the oscillation period (panel (c)) and the damping coefficient (panel (d)). Specifically, the oscillation period predicted by the axisymmetric model is only slightly larger than that predicted by the full 3D calculation, and both trends, with respect to variations of the liquid column length, are consistent with the experimental measurements.\\
\indent The damping coefficient is always larger than that computed via the axisymmetric model. This increase is attributable to three-dimensional effects, and to a slightly higher oscillation frequency. However, such an increase is bounded to less than 3\% for the lengths $l$ considered. Hence, neglecting the curved part and employing a simplified axisymmetric model appears as a justifiable assumption for the geometrical and fluid properties examined in this work. 

\bigskip
\bigskip
\begin{centering}\section{Theoretical estimate of the Stokes boundary layer contribution to the dissipation and comparison with the numerical slip-length model}\label{sec:C10_AppB}\end{centering}

In the first part of Sec.~\ref{subsec:C10_Sec2sub2}, which deals with a description of the natural properties of the system in the free-phase, we have computed numerically the damping coefficient associated with the dissipation originating in the oscillating Stokes boundary layer at the lateral wall. This numerical estimate, based on an exponentially evanescent slip-length model~\eqref{eq:C10_Slip_condition}-\eqref{eq:C10_AppB_lz2}, has provided a non-dimensional averaged damping value equal on average to $\sigma \approx 0.027$, which is less than half the one needed for a good agreement with the data ($\sigma\approx 0.6$). Such a disagreement has then motivated the introduction of an extra source of dissipation originating in the contact line region, which has eventually led to the desired value of $\sigma$.\\
\indent The use of the phenomenological contact line model~\eqref{eq:C10_free_bc_unf} and, specifically, of the chosen value of the contact line coefficient $\alpha\ne0$, has already been justified throughout the manuscript. Nevertheless, it is still worth making sure that the original numerical estimate, obtained for $\alpha=0$, represents in the first place a good prediction of the lower bound for $\sigma$, so as to not overfit the value of $\alpha$ required to increase $\sigma$ up to the desired experimental value.
\begin{figure}[]
\centering
\includegraphics[width=0.7\textwidth]{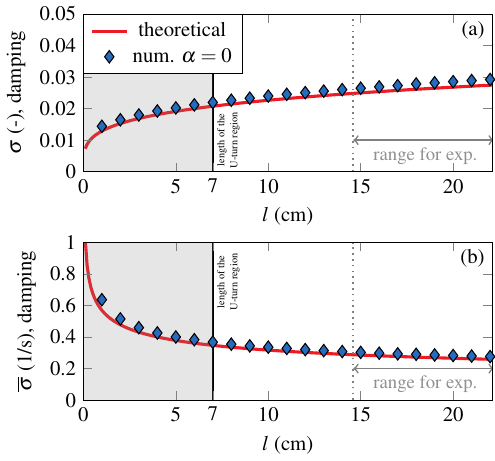}
\caption{(a) Non-dimensional, $\sigma$, and (b) dimensional, $\overline{\sigma}=\sigma\sqrt{2g/l}$, damping coefficient versus the water column length, $l\left(cm\right)$ and associated with a free contact line dynamics of the fundamental U-tube mode for $\alpha=0\,\text{rad}$. Blue diamonds: values computed fully numerical eigenvalue calculation by accounting for the variable slip length model~\eqref{eq:C10_free_bc_unf}. The red solid lines correspond to the analytical estimate of the damping coefficient as estimated in this Appendix according to equation~\eqref{eq:C10_SSP10}. The vertical black dashed lines in (a) and (b) indicate the length of the U-turn region, $\pi R\approx7\,\text{cm}$. Below this length, the liquid column is all contained in the U-turn region. In proximity and, particularly, below this limit value (as indicated by the grey-shaded regions), neglecting the curvature of the tube is no longer a justifiable assumption.}
\label{fig:C10_Fig20}
\end{figure}\\
\indent In this Appendix we therefore attempt to derive an analytical estimation of the damping coefficient produced by the Stokes boundary dissipation. To this end, as in Sec.~\ref{subsec:C10_Sec2sub2}, we neglect the tube curvature and we assume a pure free-end edge contact line condition, i.e. $\alpha=0$. Additionally, for the sake of mathematical tractability, we ignore here the curvature of the static interface, i.e. $\eta_0\left(r\right)=0$, by taking $\theta_{s}=90^{\circ}$. Note that the experimentally measured value is $\theta_{s}=80.5^{\circ}$; this angle produces a static meniscus whose characteristic length is approximately 5-6\% the tube radius, i.e. its influence is likely negligible (see Fig.~\ref{fig:C10_Fig2}).\\
\indent \textcolor{black}{Under these hypotheses, the problem of free-phase U-tube oscillations is formally equivalent to the Stokes second problem for axial oscillations governed by} 
\begin{equation}
\label{eq:C10_SSP1}
\frac{\partial w}{\partial t}=\nu\left(\frac{1}{r}\frac{\partial w}{\partial r}+\frac{\partial^2w}{\partial r^2}\right),\ \ \ \ \left.w\right|_{r=a}=W\cos{\omega_0 t},
\end{equation}
\noindent with the additional constraint the the axial velocity remains bounded for $r\rightarrow 0$. The solution of Eq.~\eqref{eq:C10_SSP1} gives the axisymmetric axial velocity profile inside the cylinder, i.e. for $0\le r\le a$,
\begin{equation}
\label{eq:C10_SSP2}
w\left(r,t\right)=W\, \text{Real}\left[\frac{I_0\left(r\sqrt{\text{i}\omega_0/\nu}\right)}{I_0\left(a\sqrt{\text{i}\omega_0/\nu}\right)}e^{\text{i}\omega_0 t}\right],
\end{equation}
\noindent where $I_0$ is the modified Bessel function of the first kind.\\
\indent We can then compute the total force exerted by the fluid on the lateral wall as
\begin{equation}
\label{eq:C10_SSP3}
F=\mu\left.\frac{\partial w}{\partial r}\right|_{r=a}=\left(\pi a l\right)\, \mu W\, \text{Real}\left[\sqrt{\frac{\text{i}\omega_0}{\nu}}\frac{I_0\left(r\sqrt{\text{i}\omega_0/\nu}\right)}{I_0\left(a\sqrt{\text{i}\omega_0/\nu}\right)}\textcolor{black}{e^{\text{i}\omega_0t}}\right],
\end{equation}
\noindent where the term $\left(\pi a l\right)$ represents the total wall surface for half tube of radius $a$ and length $l/2$. The associated power reads
\begin{equation}
\label{eq:C10_SSP4}
P=F\cdot \left.w\right|_{r=a}=\left(\pi a l\right)\, \mu W^2\, \text{Real}\left[\sqrt{\frac{\text{i}\omega_0}{\nu}}\frac{I_1\left(a\sqrt{\text{i}\omega_0/\nu}\right)}{I_0\left(a\sqrt{\text{i}\omega_0/\nu}\right)}\textcolor{black}{e^{\text{i}\omega_0t}}\right]\,\text{Real}\left[e^{\text{i}\omega_0 t}\right].
\end{equation}
\noindent The power dissipated by viscous forces during the steady-state oscillatory motion can be expressed as
\begin{equation}
\label{eq:C10_SSP5}
\langle\dot{E}\rangle=-\textcolor{black}{\frac{\omega_0}{2\pi}}\int_{0}^{\frac{2\pi}{\omega_0}}P\,\text{d}t=-\frac{\omega_0 al}{2}\, \mu W^2\, C.
\end{equation}
\noindent \textcolor{black}{with brackets $\langle\,\,.\, \,\rangle$ denoting the temporal average over one period and the auxiliary coefficient $C$ defined as}
\begin{equation}
\label{eq:C10_SSP6}
C=\int_{0}^{\frac{2\pi}{\omega_0}}\text{Real}\left[\sqrt{\frac{\text{i}\omega_0}{\nu}}\frac{I_1\left(a\sqrt{\text{i}\omega_0/\nu}\right)}{I_0\left(a\sqrt{\text{i}\omega_0/\nu}\right)}\textcolor{black}{e^{\text{i}\omega_0t}}\right]\,\text{Real}\left[e^{\text{i}\omega_0 t}\right]\,\text{d}t.
\end{equation}
\indent \textcolor{black}{Outside the thin Stokes boundary layers, the U-tube linear dynamics can be approximated by a plug flow with an interface rigidly oscillating in time at natural oscillation frequency $\omega_0^2=2g/l$ and without deforming in the radial direction.}This simple dynamics can be described by introducing the generalized coordinate $q\left(t\right)$, such that the interface position $\eta$ and the axial velocity $w$ read, respectively, $\eta=q$ and $w=\dot{q}\left(t\right)$.\\
\indent Let us now evaluate the total mechanical energy $E$, sum of the kinetic ($K$) and potential ($P$) energies, associated with the oscillatory motion:
\begin{equation}
\label{eq:C10_SSP7}
E=K+P=\frac{\rho}{2}\int_{-\frac{l}{2}}^{0}\int_{0}^{2\pi}\int_{0}^{\textcolor{black}{a}}w^2\, r\text{d}r\text{d}\phi\text{d}z\,+\,\frac{\rho g}{2}\int_{0}^{2\pi}\int_{0}^{a}\eta^2\,r\text{d}r\text{d}\phi=\frac{\rho g}{2}\pi a^2\left(\frac{\dot{q}^2}{\omega_0^2}+q^2\right).
\end{equation}
\noindent Assuming the ansatz $q\left(t\right)=D_q\left(t\right)\cos{\omega_0 t}$, one finds 
\begin{equation}
\label{eq:C10_SSP8}
E=\frac{\rho g \pi a^2}{2} \left[D_q^2+\dot{D}_q\left(\textcolor{black}{\dot{D}_q\frac{\cos^2{\omega_0 t}}{\omega_0^2}-D_q\frac{\sin{2\omega_0 t}}{\omega_0}}\right)\right]\approx \frac{\rho g \pi a^2}{2} D_q^2.
\end{equation}
\noindent with the last approximation on the right-hand side that holds for small damping, i.e. whenever $D_q\left(t\right)$ represents a slow-time damping process over the characteristic fast time-scale typical of the oscillations at frequency, \textcolor{black}{i.e. $\sim1/\omega_0$, so that $\omega_0^{-1}\dot{D}_q\ll D_q$}. The time-derivative of the total energy then reads
\begin{equation}
\label{eq:C10_SSP8bis}
\dot{E}=\rho g \pi a^2 D_q \dot{D}_q.
\end{equation}
\indent In contradistinction with the standard Stokes second problem, where the lateral wall is oscillating harmonically at a frequency $\omega_0$ with amplitude $W$, in the U-tube dynamics the sidewall is fixed and the liquid column is oscillating at frequency $\omega_0$ with amplitude $|w|=|\dot{q}|$. Recalling that $\langle\dot{E}\rangle=-\textcolor{black}{\frac{\omega_0 a l}{2}}\, \mu W^2\, C$, we can thus express $W^2$ as $|w|^2=|\dot{q}|^2=\omega_0^2 D_q^2$. Lastly, by assuming that $\langle\dot{E}\rangle\,\approx \dot{E}$,
\begin{equation}
\label{eq:C10_SSP8}
\dot{E}=\rho g \pi a^2 D_q \dot{D}_q=-\textcolor{black}{\frac{\omega_0^3 a l}{2}}\, \mu C\, D_q^2 = \langle\dot{E}\rangle\ \ \ \ \Longrightarrow\ \ \ \ \dot{D}_q=-\frac{\omega_0 \nu C}{\pi a}\, D_q,
\end{equation}
\noindent \textcolor{black}{where we have used $\omega_0^2=2g/l$}, and
\begin{equation}
\label{eq:C10_SSP9}
D_q=D_{q_0}\,\text{exp}\left(\frac{\omega_0 \nu C}{\pi a}t\right)\ \ \ \ \Longrightarrow\ \ \ \ E=\underbrace{\frac{\rho g\pi a^2}{2}D_{q_0}^2}_{E_0}\text{exp}\left(\frac{2\omega_0\nu C}{\pi a}t\right),
\end{equation}
\noindent which eventually leads to the analytical estimation of the damping coefficient $\sigma$ as
\begin{equation}
\label{eq:C10_SSP10}
\frac{E}{E_0}=\left(\frac{D_q}{D_{q_0}}\right)^2=\text{exp}\left(\frac{2\omega_0\nu C}{\pi a}t\right)=\text{exp}\left(-2\omega_0\sigma t\right)\ \ \ \ \Longrightarrow\ \ \ \ \sigma=\frac{\nu C}{\pi a}\, ,
\end{equation}
\noindent which must be compared with the numerical estimation reported in Fig.~\ref{fig:C10_FIg13}. This is done in Fig.~\ref{fig:C10_Fig20}. Both the theoretical and numerical models neglect the curvature of the tube and the extra contact line dissipation. We can see that the two predictions compare very well, hence confirming that the slip-length model~\eqref{eq:C10_Slip_condition}-\eqref{eq:C10_AppB_lz2} allows for a fair estimation of the Stokes boundary layer dissipation, as already suggested by the analysis of Bongarzone \& Gallaire (2022) \citep{bongarzone2022numerical}. This calculation also further confirms that the laminar boundary layer dissipation alone is not sufficient to justify the experimentally fitted damping coefficient.\\
\indent The effect of U-tube curvature on the damping has been discussed in Appendix~\ref{sec:C10_AppA}. The increase in the damping attributable to the three-dimensionality of the flow in the U-turn region appears too small to close to the gap with experiments, hence reinforcing the hypothesis that the additional dissipation indeed comes from the contact line dynamics.

\bibliographystyle{unsrt}
\bibliography{UTube_references}

\begin{thebibliography}{10}

\bibitem{bongarzone2021relaxation}
A.~Bongarzone, F.~Viola, and F.~Gallaire.
\newblock Relaxation of capillary-gravity waves due to contact line
  nonlinearity: A projection method.
\newblock {\em Chaos}, 31(12):123124, 2021.

\bibitem{dollet2020transition}
B.~Dollet, {\'E}.~Lorenceau, and F.~Gallaire.
\newblock Transition from exponentially damped to finite-time arrest liquid
  oscillations induced by contact line hysteresis.
\newblock {\em Phys. Rev. Lett.}, 124(10):104502, 2020.

\bibitem{ibrahim2009liquid}
R.~A. Ibrahim.
\newblock {\em Liquid sloshing dynamics: theory and applications}.
\newblock Cambridge University Press, 2005.

\bibitem{mayer2012walking}
H.~C. Mayer and R.~Krechetnikov.
\newblock Walking with coffee: Why does it spill?
\newblock {\em Phys. Rev. E}, 85(4):046117, 2012.

\bibitem{bauerlein2021phase}
B.~B{\"a}uerlein and K.~Avila.
\newblock Phase lag predicts nonlinear response maxima in liquid-sloshing
  experiments.
\newblock {\em J.~Fluid Mech.}, 925, 2021.

\bibitem{miliaiev_timokha_2023}
A.~Miliaiev and A.~Timokha.
\newblock Viscous damping of steady-state resonant sloshing in a clean
  rectangular tank.
\newblock {\em J.~Fluid Mech.}, 965:R1, 2023.

\bibitem{Lamb32}
H.~Lamb.
\newblock {\em Hydrodynamics}.
\newblock Cambridge university press, 1993.

\bibitem{Case1957}
K.~M. Case and W.~C. Parkinson.
\newblock Damping of surface waves in an incompressible liquid.
\newblock {\em J.~Fluid Mech.}, 2(2):172--184, 1957.

\bibitem{Ursell52}
F.~Ursell.
\newblock Edge waves on a sloping beach.
\newblock {\em Proc.~R.~Soc.~Lond. A}, 214:79--97, 1952.

\bibitem{Miles67}
J.~W. Miles.
\newblock Surface-wave damping in closed basins.
\newblock {\em Proc. R. Soc. A: Math. Phys. Eng. Sci.}, 297:459--475, 1967.

\bibitem{faltinsen2005liquid}
O.~M. Faltinsen and A.~N. Timokha.
\newblock {\em Sloshing}.
\newblock Cambridge University Press, 2009.

\bibitem{bongarzone2022amplitude}
A.~Bongarzone, M.~Guido, and F.~Gallaire.
\newblock An amplitude equation modelling the double-crest swirling in
  orbital-shaken cylindrical containers.
\newblock {\em J.~Fluid Mech.}, 943:A28, 2022.

\bibitem{marcotte2023super}
A.~Marcotte, F.~Gallaire, and A.~Bongarzone.
\newblock Super-harmonically resonant swirling waves in longitudinally forced
  circular cylinders.
\newblock {\em J. Fluid Mech.}, 966:A41, 2023.

\bibitem{marcotte2023swirling}
A.~Marcotte, F.~Gallaire, and A.~Bongarzone.
\newblock Swirling against the forcing: evidence of stable counter-directed
  sloshing waves in orbital-shaken reservoirs.
\newblock 2023, DOI: https://doi.org/10.48550/arXiv.2302.14579.

\bibitem{Benjamin79}
T.~B. Benjamin and J.~C. Scott.
\newblock Gravity-capillary waves with edge constraints.
\newblock {\em J.~Fluid Mech.}, 92:241--267, 1979.

\bibitem{graham1983new}
J.~Graham-Eagle.
\newblock A new method for calculating eigenvalues with applications to
  gravity-capillary waves with edge constraints.
\newblock {\em Math. Proc. Camb. Phil. Soc.}, 94(3):553--564, 1983.

\bibitem{henderson1994surface}
D.~M. Henderson and J.~W. Miles.
\newblock Surface-wave damping in a circular cylinder with a fixed contact
  line.
\newblock {\em J.~Fluid Mech.}, 275:285--299, 1994.

\bibitem{martel1998surface}
C.~Martel, J.~A. Nicolas, and J.~M. Vega.
\newblock Surface-wave damping in a brimful circular cylinder.
\newblock {\em J.~Fluid Mech.}, 360:213--228, 1998.

\bibitem{miles1998note}
J.~W. Miles and D.~M. Henderson.
\newblock A note on interior vs. boundary-layer damping of surface waves in a
  circular cylinder.
\newblock {\em J.~Fluid Mech.}, 364:319--323, 1998.

\bibitem{howell2000measurements}
D.~R. Howell, B.~Buhrow, T.~Heath, C.~McKenna, W.~Hwang, and M.~F. Schatz.
\newblock Measurements of surface-wave damping in a container.
\newblock {\em Phys. Fluids}, 12(2):322--326, 2000.

\bibitem{nicolas2002viscous}
J.~A. Nicol{\'a}s.
\newblock The viscous damping of capillary-gravity waves in a brimful circular
  cylinder.
\newblock {\em Phys. Fluids}, 14(6):1910--1919, 2002.

\bibitem{nicolas2005effects}
J.~A. Nicol{\'a}s.
\newblock Effects of static contact angles on inviscid gravity-capillary waves.
\newblock {\em Phys. Fluids}, 17(2):022101, 2005.

\bibitem{kidambi2009meniscus}
R.~Kidambi.
\newblock Meniscus effects on the frequency and damping of capillary-gravity
  waves in a brimful circular cylinder.
\newblock {\em Wave Motion}, 46(2):144--154, 2009.

\bibitem{Hocking87}
L.~M. Hocking.
\newblock The damping of capillary--gravity waves at a rigid boundary.
\newblock {\em J.~Fluid Mech.}, 179:253--266, 1987.

\bibitem{xia2018moving}
Y.~Xia and P.~H. Steen.
\newblock Moving contact-line mobility measured.
\newblock {\em J. Fluid Mech.}, 841:767--783, 2018.

\bibitem{li2019stability}
J.~Li, X.~Li, and S.~Liao.
\newblock Stability and hysteresis of faraday waves in hele-shaw cells.
\newblock {\em J. Fluid Mech.}, 871:694--716, 2019.

\bibitem{bongarzone2023revised}
A.~Bongarzone, B.~Jouron, F.~Viola, and F.~Gallaire.
\newblock A revised gap-averaged floquet analysis of faraday waves in hele-shaw
  cells.
\newblock 2023, DOI: https://doi.org/10.48550/arXiv.2306.11501.

\bibitem{blake1993dynamic}
T.~D. Blake.
\newblock Dynamic contact angle and wetting kinetics.
\newblock {\em Wettability}, 1993.

\bibitem{hamraoui2000can}
A.~Hamraoui, K.~Thuresson, T.~Nylander, and V.~Yaminsky.
\newblock Can a dynamic contact angle be understood in terms of a friction
  coefficient?
\newblock {\em J. Colloid Interface Sci.}, 226(2):199--204, 2000.

\bibitem{blake2006physics}
T.~D. Blake.
\newblock The physics of moving wetting lines.
\newblock {\em J. Colloid Interface Sci.}, 299(1):1--13, 2006.

\bibitem{navier1823memoire}
C.~L. M.~H. Navier.
\newblock M{\'e}moire sur les lois du mouvement des fluides.
\newblock {\em M{\'e}m. Acad. R. des Sci. Inst. France}, 6(1823):389--440,
  1823.

\bibitem{Keulegan59}
G.~H. Keulegan.
\newblock Energy dissipation in standing waves in rectangular basins.
\newblock {\em J.~Fluid Mech.}, 6(1):33--50, 1959.

\bibitem{Huh71}
C.~Huh and L.~E. Scriven.
\newblock Hydrodynamic model of steady movement of a solid/liquid/fluid contact
  line.
\newblock {\em J. Colloid Interface Sci.}, 35(1):85--101, 1971.

\bibitem{Davis1974}
S.~H. Davis.
\newblock On the motion of a fluid-fluid interface along a solid surface.
\newblock {\em J.~Fluid Mech.}, 65(1):71--95, 1974.

\bibitem{miles1990capillary}
J.~W. Miles.
\newblock Capillary-viscous forcing of surface waves.
\newblock {\em J.~Fluid Mech.}, 219:635--646, 1990.

\bibitem{ting1995boundary}
C.~L. Ting and M.~Perlin.
\newblock Boundary conditions in the vicinity of the contact line at a
  vertically oscillating upright plate: an experimental investigation.
\newblock {\em J.~Fluid Mech.}, 295:263--300, 1995.

\bibitem{eggers2005existence}
J.~Eggers.
\newblock Existence of receding and advancing contact lines.
\newblock {\em Phys. Fluids}, 17(8):082106, 2005.

\bibitem{Lauga2007}
E.~Lauga, M.~Brenner, and H.~Stone.
\newblock Microfluidics: the no--slip boundary condition.
\newblock {\em Springer handbook of experimental fluid mechanics}, pages
  1219--1240, 2007.

\bibitem{Eral2013}
H.~B. Eral, J.~C. M.~'T Mannetje, and J.~M. Oh.
\newblock Contact angle hysteresis: a review of fundamentals and applications.
\newblock {\em Colloid Polym.~Sci.}, 291:(2) 247--260, 2013.

\bibitem{Dussan79}
E.~B. Dussan.
\newblock On the spreading of liquids on solid surfaces: static and dynamic
  contact lines.
\newblock {\em Annu. Rev. Fluid Mech.}, 11(1):371--400, 1979.

\bibitem{rio2005boundary}
E.~Rio, A.~Daerr, B.~Andreotti, and L.~Limat.
\newblock Boundary conditions in the vicinity of a dynamic contact line:
  experimental investigation of viscous drops sliding down an inclined plane.
\newblock {\em Phys. Rev. Lett.}, 94(2):024503, 2005.

\bibitem{le2005shape}
N.~Le Grand, A.~Daerr, and L.~Limat.
\newblock Shape and motion of drops sliding down an inclined plane.
\newblock {\em J. Fluid Mech.}, 541:293--315, 2005.

\bibitem{voinov1976hydrodynamics}
O.~V. Voinov.
\newblock Hydrodynamics of wetting.
\newblock {\em Fluid Dyn.}, 11(5):714--721, 1976.

\bibitem{de1985wetting}
P.-G.~De Gennes.
\newblock Wetting: statics and dynamics.
\newblock {\em Rev. Modern Phys.}, 57(3):827, 1985.

\bibitem{cox1986dynamics}
R.~G. Cox.
\newblock The dynamics of the spreading of liquids on a solid surface. part 1.
  viscous flow.
\newblock {\em J.~Fluid Mech.}, 168:169--194, 1986.

\bibitem{snoeijer2013moving}
J.~H. Snoeijer and B.~Andreotti.
\newblock Moving contact lines: scales, regimes, and dynamical transitions.
\newblock {\em Ann. Rev. Fluid Mech.}, 45:269--292, 2013.

\bibitem{jiang2004contact}
L.~Jiang, M.~Perlin, and W.~W. Schultz.
\newblock Contact-line dynamics and damping for oscillating free surface flows.
\newblock {\em Phys. Fluids}, 16(3):748--758, 2004.

\bibitem{noblin2004vibrated}
X.~Noblin, A.~Buguin, and F.~Brochard-Wyart.
\newblock Vibrated sessile drops: Transition between pinned and mobile contact
  line oscillations.
\newblock {\em The European Physical Journal E}, 14:395--404, 2004.

\bibitem{viola2016resonance}
F.~Viola.
\newblock Resonance in swirling wakes and sloshing waves.
\newblock Technical report, EPFL, 2016.

\bibitem{Cocciaro93}
B.~Cocciaro, S.~Faetti, and C.~Festa.
\newblock Experimental investigation of capillarity effects on surface gravity
  waves: non-wetting boundary conditions.
\newblock {\em J.~Fluid Mech.}, 246:43--66, 1993.

\bibitem{amberg2022detailed}
G.~Amberg.
\newblock Detailed modelling of contact line motion in oscillatory wetting.
\newblock {\em npj Microgravity}, 8(1):1, 2022.

\bibitem{ludwicki2022contact}
J.~M. Ludwicki, V.~R. Kern, J.~McCraney, J.~B. Bostwick, S.~Daniel, and P.~H.
  Steen.
\newblock Is contact-line mobility a material parameter?
\newblock {\em npj Microgravity}, 8(1):6, 2022.

\bibitem{Viola2018a}
F.~Viola, P.-T. Brun, and F.~Gallaire.
\newblock Capillary hysteresis in sloshing dynamics: a weakly nonlinear
  analysis.
\newblock {\em J.~Fluid Mech.}, 837:788--818, 2018.

\bibitem{Viola2018b}
F.~Viola and F.~Gallaire.
\newblock Theoretical framework to analyze the combined effect of surface
  tension and viscosity on the damping rate of sloshing waves.
\newblock {\em Phys. Rev. Fluids}, 3(9):094801, 2018.

\bibitem{fiorini2022effect}
D.~Fiorini, M.~A. Mendez, A.~Simonini, J.~Steelant, and D.~Seveno.
\newblock Effect of inertia on the dynamic contact angle in oscillating
  menisci.
\newblock {\em Phys. Fluids}, 34(10):102116, 2022.

\bibitem{iguchi1982analysis}
M.~Iguchi, M.~Ohmi, and K.~Maegawa.
\newblock Analysis of free oscillating flow in a u-shaped tube.
\newblock {\em Bulletin of JSME}, 25(207):1398--1405, 1982.

\bibitem{bongarzone2022sub}
A.~Bongarzone, F.~Viola, S.~Camarri, and F.~Gallaire.
\newblock Subharmonic parametric instability in nearly brimful circular
  cylinders: a weakly nonlinear analysis.
\newblock {\em J.~Fluid Mech.}, 947:A24, 2022.

\bibitem{bongarzone2022numerical}
A.~Bongarzone and F.~Gallaire.
\newblock Numerical estimate of the viscous damping of capillary-gravity waves:
  A macroscopic depth-dependent slip-length model.
\newblock 2022, DOI: https://doi.org/10.48550/arXiv.2207.06907.

\end{thebibliography}

\end{document}